\documentclass[12pt]{article}
\usepackage{float}
\usepackage{amsmath}
\usepackage{amsfonts}
\usepackage{tcolorbox}
\usepackage{enumitem}
\usepackage{geometry}
\usepackage{graphicx} 
\usepackage{booktabs} 
\usepackage{amsthm}
\newtheorem{theorem}{Theorem}
\newtheorem{remark}{Remark}
\newtheorem{lemma}{Lemma}

\newtheorem{assumption}{Assumption}
\usepackage{tikz}
\usepackage{tabularx}
\usepackage{longtable}
\usepackage{array}
\usepackage{subcaption}
\usepackage{tcolorbox}
\usepackage{siunitx}

\usepackage[colorlinks,
            linkcolor=blue,
            anchorcolor=blue,
            citecolor=blue]{hyperref}

\newtheorem{proposition}{Proposition}
\newtheorem{corollary}{Corollary}

\usepackage[authoryear]{natbib}
\usepackage{setspace}
\title{Principled Identification of Structural Dynamic Models}

\author{\textbf{Neville Francis}$^{\mathsection}$$\quad$\textbf{Peter Reinhard
Hansen}$^{\mathsection}$\thanks{\linespread{0.9}\selectfont Corresponding author. We thank Reinhard Ellwanger, Evi Pappa, Kyle Jurado, Valerie Ramey, seminar and conference participants at the University of Pittsburgh, Seoul National University, and the Applied Time Series Econometrics Workshop at the St.~Louis Fed for constructive comments. We are especially grateful to Oriol Gonz\'alez-Casas\'us for his discussion of an earlier version of our paper at the 13th ECB Conference on Forecasting Techniques: ``Artificial intelligence in the analysis of economic narratives, forecasting, and risk assessment''.
We also thank Brent Bundick, Efrem Castelnuovo, Martin Eichenbaum, Luca Gambetti, Simon Gilchrist, Marco Lorusso, Ren\'{e}e Fry-McKibbin, and Barbara Rossi for sharing their replication files. 
Chen Tong acknowledges financial support from the Youth Fund of the National
Natural Science Foundation of China (72301227) and the Fujian Provincial Natural Science Foundation of China (2025J08008).}$\quad$\textbf{Chen Tong}$^{\ddagger}$
 \\
$^{\mathsection}${\normalsize\emph{Department of Economics, University
of North Carolina at Chapel Hill}}\\
 {\normalsize$^{\ddagger}$}{\normalsize\emph{School of Economics, Xiamen University}} }
\date{\footnotesize \today\vspace{-1cm}}

\begin{document}
\maketitle
\begin{abstract}\singlespacing
We take a new perspective on identification in structural dynamic models: rather than imposing restrictions alone, we optimize an objective. While definitive structural identification ultimately requires exogenous economic insight, a weighted correlation-maximizing objective yields an Order- and Scale-Invariant Scheme (OASIS) that selects the orthogonal rotation most aligned with designated target variables. In traditional SVARs, these targets are the reduced-form innovations, making OASIS a natural reference rotation. We show that recursive Cholesky identification is a constrained version of the same objective and that OASIS is systematically closer to perfect correlation, closing roughly twice as much of the gap as recursive orderings, both theoretically and empirically. The same framework also provides a principled estimation strategy for Proxy VARs (IV-SVARs), where the weighted criterion is essential for resolving overdetermination in multi-proxy systems while symmetrically accommodating proxy leakage. Revisiting 22 published SVARs, we find that reduced-form innovations are typically only weakly correlated, helping explain the historical robustness of recursive schemes. Applying OASIS to seminal proxy applications, however, reveals economically important leakage across shocks and shows that accounting for such leakage can materially alter substantive conclusions.
\end{abstract}

\noindent{\small\textit{Keywords:}}{\small{} Structural Vector Autoregressions, Local Projections, Proxy VARs, Cholesky decomposition.}\  
\noindent{\small\textit{{\noindent}JEL Classification:}}{\small{}
C32, C15, E00}{\small\par}

\clearpage{}

\section{Introduction}
Identifying the latent economic shocks that drive fluctuations is a central challenge in dynamic macroeconomics, whether employing structural vector autoregressions (SVARs) or local projections (LPs); \citet{Sims:1980}, \citet{Jorda:2005}, and \citet{PlagborgMollerWolf:2021}. A central assumption is that these structural shocks are mutually uncorrelated, unlike the contemporaneously correlated residuals of a reduced-form VAR. To recover structural shocks, researchers impose identification restrictions to pin down a rotation of the reduced-form innovations. Given a covariance matrix $\Sigma \in \mathbb{R}^{n \times n}$, the identification problem arises from the multiple solutions to $\Sigma = BB^\prime$. The matrix $B$ has $n^2$ elements, while the symmetric $\Sigma$ contains only $n(n+1)/2$ unique elements. Full identification therefore requires $n(n-1)/2$ restrictions. 

Traditional approaches achieve identification by imposing ``hard'' restrictions on the structural system. A standard approach is to impose a triangular structure on $B$, in which case $B$ is the Cholesky decomposition of $\Sigma$ \citep{Sims:1980, Bernanke:1986, BlanchardWatson:1986}. Other popular methods include long-run restrictions \citep{BlanchardQuah:1989, Gali:1999, FRANCIS20051379} and sign restrictions \citep{CanovaDeNicolo:2002, Uhlig:2005, RubioRamirezWaggonerZha:2010}. While each of these schemes has its own appeal, they often face criticism regarding the plausibility of the underlying exclusion assumptions \citep{CooleyLeRoy:1985, FaustLeeper:1997, FryPagan:2011, BaumeisterHamilton:2015-Ecta, Ramey:2016}. A modern alternative uses external instruments, leading to the Proxy VAR (or IV-SVAR) framework \citep{StockWatson:2012, MertensRavn:2013}. But this approach replaces internal restrictions with external ones: in its classical form, it requires each proxy to be uncorrelated with all non-target structural shocks. In multi-shock settings, these exact zero-covariance restrictions can overdetermine the system and rule out proxy leakage across shocks.

A common response to the arbitrariness of recursive identification is to report results for alternative Cholesky orderings, often under the premise that this establishes robustness or bounds the true effects. Recently, \citet{KilianPlanteRichter:2025} show that this practice can be misleading: similarity of impulse responses across recursive orderings is not evidence of correct identification, and the range spanned by alternative orderings need not bound the true structural responses. Our theoretical results help explain why. The collection of all Cholesky orderings represents a selective and narrow subset of valid structural rotations, and the data-generating process considered by \citet{KilianPlanteRichter:2025} lies far from any recursive ordering. 

In this paper, we propose a principled approach to evaluating identification: formulate the mapping between reduced-form innovations and structural shocks as an explicit optimization. Optimization-based identification has been successfully employed to isolate specific economic mechanisms, such as the maximum forecast error variance approach for identifying news or technology shocks \citep{Uhlig:2005, BarskySims:2011, FrancisOwyangRoushDiCecio:2014}. Here, we apply this principle directly to the correlation structure between structural shocks and designated target variables -- either internal VAR innovations or external proxy variables. While definitive structural identification ultimately requires exogenous economic insight, the corresponding first-order conditions can be interpreted as implicit identifying restrictions, and in special cases many familiar identification schemes can be viewed as the optimality conditions of an appropriately chosen objective. This reframes the question from ``Are the identifying restrictions plausible?'' to ``Is the objective sound for the economic question at hand?'' In short, the discourse shifts from restrictions to objective soundness.

The reduced-form covariance matrix admits many observationally equivalent structural decompositions. Without an additional criterion, these alternatives are equally admissible, even though some yield ``structural shocks'' that are economically uninformative. One such example is a purported monetary policy shock that is only weakly related to the policy rate innovation, or a narrative tax instrument that exhibits low relevance for its targeted shock. This motivates an approach that makes the intended economic association explicit and uses it to evaluate the set of admissible decompositions. For traditional SVARs, calculating the structural rotation that maximizes alignment with the designated targets provides a necessary diagnostic baseline. For Proxy VARs, the same optimization moves beyond a diagnostic role and becomes an estimation strategy for the structural system. By maximizing instrument relevance, it provides a route to point identification while naturally accommodating proxy leakage, rather than imposing exact zero-covariance restrictions that are often difficult to justify.

Our leading example is a maximum-correlation objective and variations thereof. We select the orthogonal rotation that maximizes the average correlation between each structural shock and its designated target variable, subject only to orthogonality of the structural shocks. We label the resulting rotation OASIS (Order- and Scale-Invariant Scheme) because it is invariant to the ordering and scaling of the target variables. Within the traditional SVAR framework, the target variables are the reduced-form innovations. The economic motivation is simple: each structural shock is typically associated with a particular variable. For example, a monetary policy shock is naturally associated with the federal funds rate (FFR), so we expect the structural monetary shock to be highly correlated with the reduced-form innovation to the FFR. Because reduced-form innovations generally reflect multiple structural disturbances, these correlations are nontrivial and informative about the underlying rotation.

Interestingly, this objective also connects to the statistical literature on optimal whitening. In the special case with equal weights, the OASIS rotation coincides with the ZCA-cor whitening rotation of \citet{KessyLewinStrimmer:2018}. OASIS generalizes this idea by allowing variable-specific weights and by embedding the objective within the identification problem for structural dynamic models. Unlike purely statistical identification strategies that use time-varying volatility or non-Gaussianity to identify shocks \citep{Rigobon:2003, GourierouxMonfortRenno:2017}, OASIS instead provides a moment-based approach that relies only on standard second-moment assumptions.

When applied to Proxy VARs \citep{StockWatson:2012, MertensRavn:2013, MontielOleaPaceSuriach:2021}, the target variables are the external instruments themselves\footnote{\citet{CarrieroMumtazTheodoridisTheophilopoulou:2015} show that external instruments are less susceptible to attenuation bias from measurement error than standard recursive SVARs. We extend this robustness to a broader class of imperfections, allowing proxies to suffer from systematic cross-contamination (leakage) across targeted shocks.}. Recent work also shows that departures from strict proxy exogeneity can matter empirically: \citet{AngeliniCaggianoCastelnuovoFanelli2023} show that relaxing orthogonality between a proxy and non-target shocks can substantially alter estimated multipliers, highlighting the importance of proxy contamination. In a multiple-shock setting, the classical framework requires these instruments to be strongly correlated with specific structural shocks and, in addition, imposes exact zero-covariance restrictions with non-target shocks. These additional restrictions can overdetermine the system, and \citet{AngeliniCavaliereFanelli2024} show that in multi-shock proxy-SVARs with weak proxies, point identification and standard inference generally require additional restrictions beyond those implied by the proxies alone. Because OASIS explicitly maximizes the targeted correlations, it makes instrument relevance the core estimation criterion while treating residual cross-shock correlations in a balanced way. As we show in Theorem \ref{thm:ProxyVAR}, the objective adapts naturally to this setting, recovering the structural shocks most strongly aligned with the instruments without imposing exact exclusion on the off-diagonal correlations.

Viewing identification from the perspective of an optimization problem also yields new insight into Cholesky identification. In particular, we show that Cholesky identification is closely related to the maximum-correlation objective and can be interpreted as solving a constrained version of the same problem. Specifically, a Cholesky decomposition can be understood as the solution to a sequence of optimization problems: in each step, the objective is to maximize the correlation between a structural shock and its corresponding reduced-form innovation, subject to an increasing number of orthogonality constraints. Although the resulting rotation depends on the ordering of variables, all Cholesky decompositions share this common objective and differ only in the recursive constraints imposed. This shared implicit objective helps explain why alternative Cholesky orderings often yield similar results.

Although applied SVAR papers rarely state high correlation with target innovations as an explicit goal, it is nevertheless a pervasive empirical regularity in the literature. In sixteen of the twenty-two SVAR studies we examine, the average correlation between Cholesky-identified structural shocks and their reduced-form targets exceeds 90\%; across the full set of studies, this average ranges from 78.5\% to 99.8\%. 

Order sensitivity induced by triangular factorizations arises in other ways in VARs. For instance, the connectedness model by \citet{DieboldYilmaz:2009} used Cholesky to define variable-specific shocks before adopting the generalized impulse-response/variance-decomposition framework of \citet{KoopPesaranPotter:1996-JoE} and \citet{PesaranShin:1998}. The ``generalized'' shocks are order- and scale-invariant, but are not uncorrelated. \citet{ChanKoopYu:2024-JBES} address a related but distinct aspect of order-invariance and obtain an \emph{order-invariant likelihood/posterior} for large Bayesian VARs by abandoning triangularity and exploiting stochastic volatility for identification (unique up to signs/permutations). Their decomposition focuses on the (time-varying) reduced-form covariance matrix. This is fundamentally different from OASIS, which yields an orthogonal shock system that is order- and scale-invariant. 

In the traditional SVAR setting, let $\bar\rho_\ast$ and $\bar\rho_\mathrm{c}$ denote the average correlation between structural shocks and their reduced-form targets under OASIS and Cholesky, respectively. Because OASIS explicitly maximizes this average correlation, we necessarily have $\bar\rho_\ast \geq \bar\rho_\mathrm{c}$. Moreover, we show, theoretically and empirically, that OASIS closes roughly twice as much of the gap to perfect correlation ($\bar\rho =1$) as recursive Cholesky identification.

A second theoretical result concerns recursive Cholesky identification: although the resulting structural rotation and impulse responses depend on the variable ordering, all recursive Cholesky decompositions yield very similar values of $\bar\rho_\mathrm{c}$. Hence, the choice of ordering primarily determines how the total correlation is distributed across shocks, rather than the overall level of alignment.

Furthermore, we show that the average correlations, $\bar\rho_\ast$ and $\bar\rho_\mathrm{c}$, decline as the reduced-form innovations become more strongly contemporaneously correlated. Across the 22 empirical SVAR studies we revisit, the correlation matrix of the reduced-form errors is typically close to the identity matrix. This weak contemporaneous correlation helps explain why historical applications of recursive identification have often produced high average correlations and relatively stable results across orderings. These empirical findings are closely aligned with our theoretical predictions.

The remainder of the paper is organized as follows. Section \ref{sec:Identification} formally introduces the OASIS framework and derives its core theoretical properties, alongside comparative theoretical results for recursive Cholesky identification. Section \ref{sec:ProxyVAR} extends the OASIS methodology to Proxy VARs, demonstrating how the framework delivers a principled route to point identification by symmetrically accommodating proxy leakage. Section \ref{sec:Studies} revisits 22 prominent empirical SVAR studies to evaluate the apparent historical robustness of Cholesky orderings in light of our theoretical results. Section \ref{sec:ProxyApplications} then applies OASIS to seminal Proxy VAR models, including the macroeconomic multiplier analysis of \citet{MertensRavn:2013} and the financial shock model of \citet{StockWatson:2012}, illustrating how accounting for proxy leakage can materially alter substantive conclusions. Finally, Section \ref{sec:Summary} offers concluding remarks. All mathematical proofs are relegated to Appendix A.

\section{Identification Schemes}\label{sec:Identification}

Identification is central to empirical macroeconomics, from structural VARs to local projections (LPs) and IRF matching. To establish a common framework for discussing different identification schemes, we first set out the notation and assumptions that underlie these models.

We begin by formalizing the basic environment shared by these methods, and we largely follow the notation used in \citet{Hamilton:1994}. Let $\varepsilon\in\mathbb{R}^n$ denote a vector of reduced-form shocks with a general nonsingular covariance matrix, $\Sigma_{\varepsilon\varepsilon}\equiv\operatorname{var}(\varepsilon)$, and let $u=A^\prime \varepsilon$ denote a vector of structural shocks for some matrix $A\in\mathbb{R}^{n\times n}$. These are time series, but to simplify the exposition, we suppress the subscript-$t$ notation in this subsection.

The structural shocks, $u=A^\prime\varepsilon$, are assumed to be uncorrelated and normalized; $\operatorname{var}(u)=I_n$, hence the requirement is $A^\prime\Sigma_{\varepsilon\varepsilon} A=I_n$.
The set of $A$-matrices satisfying this requirement is denoted $$\mathcal{A}\equiv\{A\in\mathbb{R}^{n\times n}:A^\prime\Sigma_{\varepsilon\varepsilon} A=I_n\},$$ and is uncountable 
for any nonsingular covariance matrix, $\Sigma_{\varepsilon\varepsilon}$. 
The (lower triangular) Cholesky decomposition of the covariance matrix, $\Sigma_{\varepsilon\varepsilon}$, is a common choice, which is given by $A_\mathrm{c}^\prime =L^{-1}$, where $L$ is the lower triangular matrix satisfying $LL^\prime=\Sigma_{\varepsilon\varepsilon}$.
This particular choice for $A$ can be characterized as a sequential optimization problem. 
\begin{proposition}[Cholesky characterization]\label{prop:Chol} Consider the sequence of optimization problems, 
\begin{equation}\label{eq:CholRecOptimize}
a_j \equiv \arg\max_{a:a^\prime \Sigma_{\varepsilon\varepsilon} a=1} \operatorname{corr}(a^\prime \varepsilon, \varepsilon_j), \quad \text{subject to}\quad a^\prime \Sigma_{\varepsilon\varepsilon} a_i=0,\ \forall i<j,
\end{equation}
for $j=1,\ldots,n$. Then $a_j$ is the $j$-th column of $A_\mathrm{c}$.
\end{proposition}
It follows that identification based on the Cholesky decomposition ensures a relatively high average correlation between the structural shocks $u_1,\ldots,u_n$ and their corresponding reduced-form shocks $\varepsilon_1,\ldots,\varepsilon_n$.
This seems reasonable, because the structural shock to the $j$-th variable would naturally be embodied in $\varepsilon_j$. Naturally, $\varepsilon_j$ is also contaminated with other shocks (except for $j=1$), and this is responsible for $\Sigma_{\varepsilon\varepsilon}$ being a non-diagonal covariance matrix. 

The Cholesky representation depends on the ordering of variables, and the chosen ordering is sometimes criticized for being (partly) arbitrary. The first structural shock will always be perfectly correlated with the corresponding reduced-form shock and, as $j$ increases, the correlation between structural shocks and the corresponding reduced-form shocks tends to decrease because the number of constraints in (\ref{eq:CholRecOptimize}) increases with $j$. It can therefore be said that Cholesky prioritizes high correlation for shocks associated with variables that appear first in the system.

An important point of this paper is that the basic principle that underlies Cholesky identification does not require a variable ordering. It is possible to treat all dimensions equally and maximize the total (or average) correlation. This maximization can be done simultaneously over all dimensions, rather than the constrained sequential maximization implicit in Cholesky. This leads to OASIS that we will introduce next.

\subsection{A Maximum Correlation Objective}
Motivated by the implicit objective of Cholesky identification, we can simply 
maximize the average correlation between the elements of $u=A^\prime\varepsilon$
and the corresponding elements of $\varepsilon$.
In other words, we can drop the constraint imposed by the recursive ordering used in a Cholesky decomposition and simply maximize the average correlation, $\bar\rho(A)=\tfrac{1}{n}\sum_{i=1}^{n}\operatorname{corr}(u_{i},\varepsilon_{i})$.  
This defines an order- and scale-invariant reference rotation, which we label as OASIS (Order- and Scale-Invariant Scheme). 

More generally, we can maximize the weighted-correlation criterion
$$\rho_w(A)=\sum_{i=1}^{n}w_{i}\operatorname{corr}(u_{i},\varepsilon_{i}),\qquad\text{for } A\in\mathcal{A},$$ for positive weights, $w_i>0$, for $i=1,\ldots,n$, where $u=A^\prime\varepsilon$. 

We introduce the following notation. Let $C_{\varepsilon\varepsilon}\equiv \operatorname{corr}  (\varepsilon)$ denote the correlation matrix of the reduced-form shocks. Then
$$C_{\varepsilon\varepsilon}=\Lambda_{\sigma_\varepsilon}^{-1}\Sigma_{\varepsilon\varepsilon} \Lambda_{\sigma_\varepsilon}^{-1}, 
$$
where $\Lambda_{\sigma_\varepsilon}\equiv\operatorname{diag}(\sigma_1,\ldots,\sigma_n)$ 
is the diagonal matrix of standard deviations, such that $\sigma^2_i=\mathrm{var}(\varepsilon_{i})$, $i=1,\ldots,n$ are the diagonal elements of $\Sigma_{\varepsilon\varepsilon}$. 
Analogously, we define the weighting matrix, 
$\Lambda_w=\operatorname{diag}(w_{1},\ldots, w_{n})$, and let $$\lambda_{w,1},\ldots,\lambda_{w,n} \quad\text{denote the eigenvalues of}\quad\Lambda_w C_{\varepsilon\varepsilon}\Lambda_w.$$
Moreover, for a symmetric matrix, $M$,  we let $M^{1/2}$ denote the symmetric square root of $M$, and if, in addition, $M$ is positive definite, then $M^{-1/2}$  is defined by the inverse of $M^{1/2}$.\footnote{The symmetric square root is given from the eigendecomposition. Note that $(\Lambda_w C_{\varepsilon\varepsilon}\Lambda_w)^{-1/2}\neq \Lambda_w^{-1/2}C_{\varepsilon\varepsilon}^{-1/2}\Lambda_w^{-1/2}$, unless $\Lambda_w$ and $C_{\varepsilon\varepsilon}$ commute.} 
\begin{theorem}[Weighted OASIS]\label{thm:WeightedOASIS} Suppose $\det(\Sigma_{\varepsilon\varepsilon})>0$ and $w_i>0$ $i=1,\ldots,n$. Then 
\[
\rho_w(A_\ast)=\sum_{i=1}^{n}\lambda_{w,i}^{1/2}\geq \rho_w(A),\quad\text{ for all }A\in\mathcal{A},
\]
where $A_\ast=\Lambda_{\sigma_\varepsilon}^{-1}\Lambda_w(\Lambda_w C_{\varepsilon\varepsilon}\Lambda_w)^{-1/2}$ is the unique maximizer. This rotation is order- and scale-invariant.
\end{theorem}

Note $A_\ast^\prime \Sigma_{\varepsilon\varepsilon} A_\ast = I_n$ by construction, since $(\Lambda_w C_{\varepsilon\varepsilon} \Lambda_w)^{-1/2}$ is the symmetric inverse square-root of $\Lambda_w C_{\varepsilon\varepsilon} \Lambda_w$.

The weighted maximum correlation criterion allows the researcher to tilt the baseline rotation toward more reliable or the most policy-relevant shocks by assigning variable-specific weights $w_i>0$ in the objective $\rho_w(A)=\sum_i w_i\operatorname{corr}(u_i,\varepsilon_i)$. In practice, one may downweight dimensions believed to suffer greater measurement error (e.g., set $w_i$ proportional to a reliability metric such as $1/\sigma_i$, where $\sigma^2_i=\operatorname{var}(\varepsilon_i)$), or upweight variables whose shocks are of primary interest. The resulting solution is order- and scale-invariant but concentrates correlation where the signal is stronger or where the research question warrants greater emphasis.

\subsubsection{Equal Weights}
The special case with equal weights, $w_1=\cdots=w_n$, we have $\Lambda_w = cI_n$, so the eigenvalues of $\Lambda_w C_{\varepsilon\varepsilon} \Lambda_w$ are proportional to the eigenvalues of $C_{\varepsilon\varepsilon}$, denoted $\lambda_1,\ldots,\lambda_n$. Specifically,
$$
\lambda_i(w)=c^2\lambda_i, \qquad i=1,\ldots,n.
$$
Because multiplying all weights by the same positive constant only rescales the objective and does not affect its maximizer, we normalize to $w_1=\cdots=w_n=1/n$ and define
$$
\bar\rho(A)\equiv\frac{1}{n}\sum_{i=1}^{n}\operatorname{corr}(u_i,\varepsilon_i).
$$
\begin{corollary}[OASIS]\label{cor:OASIS} 
Suppose that $\det(\Sigma_{\varepsilon\varepsilon})>0$. Then$$\bar\rho(A_\ast)=\frac{1}{n}\sum_{i=1}^{n}\lambda_{i}^{1/2}\geq \bar\rho(A)\quad\text{ for all }A\in\mathcal{A},$$where $A_\ast=\Lambda_{\sigma_\varepsilon}^{-1}C_{\varepsilon\varepsilon}^{-1/2}$ is the unique solution to $\max_{A\in \mathcal{A}}\bar\rho(A)$.
So, $u^\ast=A_\ast^\prime \varepsilon$ attains the maximum average correlation, $\frac{1}{n}\sum_{i=1}^{n}\operatorname{corr}(u^\ast_{i},\varepsilon_{i})=\frac{1}{n}\sum_{i=1}^{n}\lambda_{i}^{1/2}$.
\end{corollary}
Unlike the eigendecomposition of $\Sigma_{\varepsilon\varepsilon}$, which is not scale-invariant, OASIS uses the eigendecomposition of the correlation matrix $C_{\varepsilon\varepsilon}=\Lambda_{\sigma_\varepsilon}^{-1}\Sigma_{\varepsilon\varepsilon}\Lambda_{\sigma_\varepsilon}^{-1}$, yielding $A^{\ast}=\Lambda_{\sigma_\varepsilon}^{-1}C_{\varepsilon\varepsilon}^{-1/2}$ that is invariant to both scale and ordering.
Here $C_{\varepsilon\varepsilon}^{1/2}=Q\Lambda_\lambda^{1/2}Q^{\prime}$ and $C_{\varepsilon\varepsilon}^{-1/2}=Q\Lambda_\lambda^{-1/2}Q^{\prime}$, where 
$C_{\varepsilon\varepsilon}=Q\Lambda_\lambda Q^{\prime}$ is the 
eigendecomposition of $C_{\varepsilon\varepsilon}$, i.e. $Q^{\prime}Q=I_n$ and
$\Lambda_\lambda\equiv\operatorname{diag}(\lambda_1,\ldots,\lambda_n)$.
\footnote{This equal-weight problem is considered in \citet{HansenTong:2026}, where a vector of correlated standardized returns, $z = Bu$, is represented as a linear transformation of a vector of uncorrelated standardized variables, $u$. For interpretability in these models, it is desirable that each component $z_i$ be highly correlated with its corresponding $u_i$, and the symmetric square root of the correlation matrix, $B = C_{\varepsilon\varepsilon}^{1/2}$, solves this problem.
The solution to this problem is known as  Zero-phase Component Analysis whitening, see \citet{KessyLewinStrimmer:2018}.}

A key property, which we used in the proofs, is that any $A\in\mathcal{A}$ can be expressed as 
$A=A_\ast R$, where $R$ is orthonormal (a rotation matrix). That
 $R$ is orthonormal follows by $I_n=A^\prime\Sigma_{\varepsilon\varepsilon} A=R^\prime A_\ast^\prime\Sigma_{\varepsilon\varepsilon} A_\ast R=R^\prime R$, and we have
\begin{equation}R=A_\ast^{-1} A = ( A^{-1}A_\ast)^\prime,\label{eq:Rotation}
\end{equation}
where the last identity follows by $A=A_\ast R\Leftrightarrow AR^\prime=A_\ast RR^\prime\Leftrightarrow R^\prime=A^{-1}A_\ast $.
Moreover, for two identification schemes, $A_1$ and $A_2$ say, the rotation matrix, $R_{1,2}=A_1^{-1}A_2$, characterizes how $u_1= A_1^\prime \varepsilon$ can be rotated into $u_2= A_2^\prime \varepsilon = R^\prime A_1^\prime \varepsilon =R^\prime u_1$.
This helps explain differences in impulse response functions for different identification schemes. We will make use of this in our empirical analysis in Section \ref{sec:LSZ96detailed}.

\subsection{Applications to SVARs and Local Projections}

Let $\{X_t\}$ be a time series and let $\mathcal{F}_t$ be a filtration to which $X_t$ is adapted. We are interested in how a shock to an element of $X_t$ propagates to future values of $X_t$. 
In SVARs and LPs, the reduced-form shocks are typically defined by a linear projection:
$$\varepsilon_t = X_t - \Gamma Z_{t-1},$$
where $Z_t\in\mathcal{F}_t$. In a vector autoregression of order $p$, 
$\Phi(L)X_t=\mu+\varepsilon_t$, we have $\Gamma=(\mu,\Phi_{1},\ldots,\Phi_{p})$ and $Z_{t-1}=(1,X_{t-1},\ldots,X_{t-p})^\prime$. More generally, $Z_{t-1}$ can include lagged values of other variables.

The impulse response function is defined by:
$$\operatorname{IRF}(h)=\operatorname{cov}(X_{t+h},u_{t}),$$
where the $(i,j)$-th element of $\operatorname{IRF}(h)$ measures the linear impact of $u_{j,t}$ on $X_{i,t+h}$.

In an SVAR, we have $\varepsilon_t=Bu_t$ with $B^\prime=A^{-1}$, and if the underlying VAR is invertible, then $\operatorname{IRF}(h) = \Psi_h B$, where $\Psi_h$ is the $h$-th coefficient matrix in the moving-average (MA) representation: 
\begin{equation}
    X_t = \Phi(1)\mu + \sum_{h=0}^\infty \Psi_h \varepsilon_{t-h},
\end{equation}
because $\Psi_h \varepsilon_{t-h}=\Psi_h (A^\prime)^{-1}A^\prime \varepsilon_{t-h}=\Psi_h B u_{t-h}$.

A local projection does not recover the IRF by deducing the MA coefficients from the VAR. Instead, the IRF is obtained from regressions, such as $X_{t+h}=\mu_h + \Theta_h \varepsilon_t+e_{t,t+h}$ for $h=0,1,2,\ldots$, and the IRF (in matrix form) is given by $\operatorname{IRF}(h) = \Theta_h A^{-1\prime}= \Theta_h B$. Alternatively, we can regress $X_{t+h}$ on $u_t=A^\prime\varepsilon_t$ and a constant, in which case the IRF is simply the coefficient matrix on $u_t$.

It is simple to compute the IRF for a specific economic identification scheme from those of a reference rotation. The relation between any IRF and that of the OASIS baseline is the following.

\begin{proposition}\label{prop:IRF}
    Let $\operatorname{IRF}(h)$ denote the IRF resulting from identification with $A\in\mathcal{A}$. Then
    $$\operatorname{IRF}(h)=\operatorname{IRF}^\ast(h)R,$$ 
    where $\operatorname{IRF}^\ast(h)$ is the IRF for $A_\ast$ (the OASIS baseline) and $R=A_\ast^{-1}A$ is a rotation (orthonormal) matrix. 
\end{proposition}

Discrepancies in the IRFs can be diagnosed using the rotation matrix, $R$, that shows how structural shocks from a specific economic identification can be expressed as a rotation (orthonormal linear combination) of the baseline OASIS shocks, which may explain differences between IRFs across identification schemes. It is worth noting that two IRFs can be different even if their underlying structural shocks are highly correlated with each other and with the reduced-form shocks.

\subsection{OASIS: Twice the Proximity to Perfect Correlation}

Early in our empirical analysis, we noticed that
$$(1-\bar{\rho}_{\mathrm{c}})\approx 2 ( 1-\bar{\rho}_\ast),$$
where $\bar{\rho}_{\mathrm{c}} \equiv \bar{\rho}(A^{\mathrm{c}})$ and $\bar{\rho}_\ast\equiv \bar{\rho}(A_\ast)$.
The empirical ratio $(1-\bar{\rho}_{\mathrm{c}})/( 1-\bar{\rho}_\ast)$ ranged from 1.83 to 2.22 across all studies, such that the average correlation under OASIS is about half as far from unity as that achieved by Cholesky. This is no coincidence, as the following Theorem shows. 
\begin{theorem}\label{thm:Proximity}
Let $d(C_{\varepsilon\varepsilon})=\frac{1}{n}\sum_{i\neq j}C_{\varepsilon\varepsilon,ij}^{2}=\frac{1}{n}\left\Vert C_{\varepsilon\varepsilon}-I_n\right\Vert _{F}^{2}>0$. Then 
\begin{eqnarray}
\bar{\rho}_\ast & = & 1-\tfrac{1}{8}d(C_{\varepsilon\varepsilon})+\tfrac{1}{n}O(\operatorname{tr}\{E^{3}\}), \label{eq:ApproxOASIS}\\
\bar{\rho}_{\mathrm{c}} & = & 1-\tfrac{1}{4}d(C_{\varepsilon\varepsilon})+\tfrac{1}{n}O(\operatorname{tr}\{E^{3}\}),\label{eq:ApproxCholesky}
\end{eqnarray}
and $(1-\bar{\rho}_{\mathrm{c}})/(1-\bar{\rho}_\ast)=2(1+O(\Vert E\Vert_F))=2\left(1+O(\sqrt{d(C_{\varepsilon\varepsilon})})\right)$ where $E=C_{\varepsilon\varepsilon}-I_n$.
\end{theorem}
For the special case where $C_{\varepsilon\varepsilon}$ is an equicorrelation matrix, $C_{\varepsilon\varepsilon,ij}=\rho$
for all $i\neq j$, we have $d(C_{\varepsilon\varepsilon})=\frac{1}{n}\sum_{i\neq j}\rho^{2}=(n-1)\rho^{2}$. Thus, we should expect the average correlation between structural shocks and reduced-form shocks to decrease with the dimension, $n$. The exact result for the equicorrelation case is:
\begin{corollary}\label{cor:Equicorr}
Suppose that $C_{\varepsilon\varepsilon}$ is an equicorrelation matrix, such that $[C_{\varepsilon\varepsilon}]_{ij}=\rho\in(-\tfrac{1}{n-1},1)$
for all $i\neq j$. Then
\begin{align}
\bar{\rho}_\ast & = \tfrac{1}{n}\sqrt{1+(n-1)\rho}+\sqrt{1-\rho}\left(1-\tfrac{1}{n}\right),\label{eq:EquiCorrOasis}\\
\bar{\rho}_{\mathrm{c}} & = \tfrac{1}{n}\sum_{k=1}^{n}\sqrt{1-\tfrac{(k-1)\rho^{2}}{(k-2)\rho+1}}.\label{eq:EquiCorrChol}
\end{align}
\end{corollary}

Theorem \ref{thm:Proximity} shows that OASIS (relative to Cholesky) reduces the distance
between structural shocks and reduced-form shocks by a factor of two.
This adds another testable implication that we will explore in the empirical section.

Note that $d(C_{\varepsilon\varepsilon})$ is $(n-1)$ times the average squared correlation coefficient, $\frac{1}{n(n-1)}\sum_{i\neq j}C_{\varepsilon\varepsilon,ij}^{2}$, from which it follows that $d(C_{\varepsilon\varepsilon})=\frac{1}{n}\left\| C_{\varepsilon\varepsilon}-I_n \right\| _{F}^{2}$, where $\left\Vert \cdot \right\Vert _{F}$ is the  Frobenius norm.

Another significant consequence of Theorem~\ref{thm:Proximity} concerns the role of variable ordering in Cholesky identification. 
\begin{corollary}\label{cor:CholeskySimilar}
      The average correlation between structural shocks and reduced-form shocks, in a Cholesky decomposition, satisfies
      $\bar\rho_\mathrm{c}=1-d(C_{\varepsilon\varepsilon})/4+\tfrac{1}{n}O(\operatorname{tr}\{E^{3}\})$, regardless of the chosen ordering of the variables.
\end{corollary}
Thus, the choice of variable ordering in Cholesky identification has little effect on the average correlation between structural and reduced-form shocks, especially when the reduced-form shocks are only moderately or weakly correlated. Low correlation between reduced-form shocks is found in nearly all of the empirical studies we revisit. This result can, in part, explain why many Cholesky-based empirical findings are reported to be robust to the ordering of variables.

\subsection{Observational Equivalence}

A natural question is whether OASIS recovers a particular structural rotation in a simulation where the data are generated from a pre-specified matrix, such as a Cholesky factor. This is not the relevant benchmark. A simulation may hard-code a latent rotation, but if estimation uses only the reduced-form covariance matrix, that rotation is not identified from second moments alone. Many structural decompositions are observationally equivalent in the sense that they imply the same reduced-form covariance matrix. Consequently, unless the same identifying restrictions used to construct the DGP are also imposed in estimation, no statistical procedure can be expected to recover that particular rotation in population. OASIS solves a different problem: it selects the unique order- and scale-invariant rotation that maximizes the average correlation between structural shocks and their corresponding reduced-form innovations. 
Thus, if the DGP is generated from a lower-triangular matrix, OASIS will generally not recover that matrix, not because it fails, but because the true Cholesky matrix is almost never the optimal solution to the maximum-correlation objective.

\section{OASIS for Proxy VARs}\label{sec:ProxyVAR}

Proxy VARs rely on external measures to identify structural shocks (for a recent comprehensive review of this methodology, see \citet{BrunsLutkepohl:2026}), and the maximum-correlation criterion provides a principled way to estimate the rotation most aligned with the instruments. Let $z\in\mathbb{R}^r$ denote a vector of narrative/external instruments, where we suppress the time index. 

As in an SVAR, the structural shocks are given by $A^{\prime}\varepsilon\in\mathbb{R}^n$, but these are partitioned into $(u^\prime,v^\prime)^\prime=A^{\prime}\varepsilon$, where the elements of $u=\boldsymbol{a}^{\prime}\varepsilon\in\mathbb{R}^r$ are those identified from the elements in $z$, and $v=\boldsymbol{b}^\prime\varepsilon$ is a vector of auxiliary shocks. Here $\boldsymbol{a}$ collects the $r$ identified directions and plays the same role for the instrumented shocks that $A$ did earlier for the full system, and the objective is to maximize 
$$
g(\boldsymbol{a})=\sum_{j=1}^r w_j \operatorname{corr}(u_j,z_j),\qquad\text{for}\quad\boldsymbol{a}\in \mathcal{A}_r=\{\boldsymbol{a}\in\mathbb{R}^{n \times r}: \boldsymbol{a}^\prime\Sigma_{\varepsilon\varepsilon} \boldsymbol{a} = I_r\}.
$$
The essential relevance condition is that $z_j$ is correlated with the $j$-th targeted structural shock, $u_j$, for $j=1,\ldots,r$.

The standard approach for establishing the validity of these external proxies is to impose a strict exogeneity condition, a framework formally bridged to SVARs by \citet{StockWatson:2012}. Under their foundational definition, a valid proxy must be relevant to its targeted shock and strictly orthogonal to all other structural shocks in the system. For identifying a single structural shock ($r=1$), this strict orthogonality condition holds exactly. A conceptual difficulty arises when the classical one-to-one exclusion pattern is imposed in multi-proxy settings with $r \ge 2$. In that formulation, each instrument $z_i$ must be orthogonal both to the auxiliary shocks $v$ and to all other targeted shocks $u_j$ for $j \ne i$. Although this may be true in population, the empirical cross-correlation matrix will almost surely fail to exhibit this exact sparse structure, because the implied zero restrictions outnumber the effective rotational degrees of freedom. More importantly, this diagonal-plus-zero specification is stronger than is needed for structural identification. As shown in Section 3.2, identification only requires sufficient relevance and rank conditions, not exact exclusion of all off-diagonal instrument-shock correlations.

Specifically, achieving this sparse structure requires the rotation to zero out $r(r-1)+r(n-r)=r(n-1)$ elements of the empirical matrix $C_{\varepsilon z}$. After accounting for the orthogonality constraint and the irrelevance of rotations within the $(n-r)$-dimensional auxiliary shock space, only $nr-r(r+1)/2$ effective degrees of freedom remain. The difference is exactly $r(r-1)/2$, so the system is exactly identified only when $r=1$ and becomes increasingly overdetermined as the number of proxies grows. Thus, for $r\geq 2$, the exact diagonal-plus-zero exclusion structure is generically overdetermined in empirical applications. This mathematical friction explains why researchers attempting to identify multiple shocks simultaneously are often forced to rely on arbitrary recursive orderings (Cholesky decompositions) to resolve the exact shortage of degrees of freedom.

More generally, the diagonal structure imposed on the $r \times r$ target block is stronger than is needed for system identification. In many multi-proxy settings, it is sufficient that the correlation between the instruments and the targeted shocks has full rank, rather than being diagonal. Crucially, exact exclusion restrictions are not required for identification under the OASIS framework; relevance alone provides sufficient structure to determine the identified shock space through a system-wide objective. Rather than forcing a potentially impossible set of zero-covariance restrictions on $A^\prime C_{\varepsilon z}$, OASIS treats the instruments strictly as target directions. By maximizing the aligned correlations, it naturally absorbs the empirical cross-correlations into the unconstrained off-diagonal elements. This accommodates the reality of noisy, cross-correlated proxies by balancing these off-diagonal associations according to the relevance objective. Furthermore, while traditional Proxy VAR frameworks require Two-Stage Least Squares or GMM weighting matrices to handle overidentified systems where there are multiple instruments for a given structural shock, the OASIS framework directly handles this case. The singular value decomposition of the cross-correlation matrix extracts the optimal linear combination of the instruments that maximizes the relevance objective, thereby avoiding the need to specify an additional weighting matrix.

For this problem, we introduce $C_{\varepsilon\varepsilon}=\operatorname{corr}(\varepsilon)$, $C_{\varepsilon z}=\operatorname{corr}(\varepsilon,z)$, and $\Lambda_w=\operatorname{diag}(w_1,\ldots,w_r)$, and a key quantity is
$$
\Xi = C_{\varepsilon \varepsilon}^{-1/2}C_{\varepsilon z}\Lambda_w\in\mathbb{R}^{n\times r},
$$ 
and its singular value decomposition (SVD), $\Xi=U\Lambda_\xi V^\prime$, for which $U^\prime U=V^\prime V=I_r$ and $\Lambda_\xi=\operatorname{diag}(\xi_1,\ldots,\xi_r)$. 

\begin{theorem}\label{thm:ProxyVAR}
    Suppose $\det(\Sigma_{\varepsilon\varepsilon})>0$ and $w_j > 0$ for all $j=1,\ldots,r$. Then   
$$
g(\boldsymbol{a})\leq g(\boldsymbol{a}_\ast)=\sum_{i=1}^r \xi_i,\quad\forall \boldsymbol{a}\in\mathcal{A}_r,
$$
where $\boldsymbol{a}_\ast\equiv \Lambda_{\sigma_\varepsilon}^{-1}C_{\varepsilon\varepsilon}^{-1/2}UV^\prime$. If $C_{\varepsilon z}\in\mathbb{R}^{n\times r}$ has full column rank $r$ then $\boldsymbol{a}_\ast$ is the unique maximum of $g$ over $\mathcal{A}_r$.
\end{theorem}

Geometrically, Theorem \ref{thm:ProxyVAR} demonstrates that after whitening the reduced-form innovations via $C_{\varepsilon\varepsilon}^{-1/2}$, the OASIS objective simplifies to maximizing the trace of the scaled cross-correlation matrix against an orthonormal basis. This mathematical structure is a well-known (orthogonal Procrustes problem) with an exact analytical solution given from the singular value decomposition of $\Xi$. 

Beyond identifying the unique maximum, the geometry of the optimal rotation $\boldsymbol{a}_\ast$ characterizes the relationship between the structural shocks and the external instruments. Specifically, the resulting vector of structural shocks satisfies the following three first-order conditions:

\begin{corollary}\label{cor:ProxyVarFOC}
Given the assumptions of Theorem \ref{thm:ProxyVAR}. Let $u^\ast_t = \boldsymbol{a}_\ast^\prime \varepsilon_t$. Then $(i)$ $\operatorname{var}(u^\ast_t)=I_r$, $(ii)$ 
\begin{equation}\label{eq:GMM2pop}
    (I_n- C_{\varepsilon\varepsilon} \Lambda_{\sigma_\varepsilon} \boldsymbol{a}_\ast \boldsymbol{a}_\ast^\prime \Lambda_{\sigma_\varepsilon}) C_{\varepsilon z}=0,
\end{equation}
and $(iii)$ $\operatorname{corr}(u^\ast_t,z_t)\Lambda_w$ is symmetric.
\end{corollary}

These properties offer a clear econometric interpretation. Property $(ii)$ states that the component of the reduced-form shocks orthogonal to the identified shock space is strictly orthogonal to the instruments. Property $(iii)$ provides a geometric resolution to the multi-shock dilemma: rather than forcing the off-diagonal elements of the cross-correlation matrix to be exactly zero, the optimally weighted instrument-shock correlation matrix is instead symmetrized. This mathematically accommodates collinear macroeconomic proxies without imposing exact zero-restrictions.

Interestingly, in the special case with equal weights ($\Lambda_w=I_r$), the singular values $\xi_1,\ldots,\xi_r$ of $\Xi=C_{\varepsilon\varepsilon}^{-1/2}C_{\varepsilon z}$ provide a natural diagnostic for instrument relevance under the OASIS criterion. Because the maximized objective is exactly the sum of these singular values, each $\xi_i$ quantifies the identifying strength along a distinct, orthogonal dimension of the instrument space. A large singular value indicates robust identifying content, whereas a small $\xi_i$ signals weak identification, warning the researcher that at least one direction in the structural subspace is poorly pinned down by the data. 

\subsection{Structural Identification by OASIS}

In this section, we establish conditions under which OASIS identifies the true structural shocks, $u=\boldsymbol{a}_0^\prime \varepsilon \in \mathbb{R}^{r}$, $\boldsymbol{a}_0\in\mathcal{A}_r$, from a vector of instruments $z\in\mathbb{R}^r$. Throughout, we assume that the second moments of shocks and instrumental variables are finite, and from standard linear projection arguments, we have: 
\begin{equation}\label{eq:localproject}
    z = \mu_z + \Sigma_{z\varepsilon}\Sigma_{\varepsilon\varepsilon}^{-1} \varepsilon + \eta,\qquad \mu_z\equiv\mathbb{E}[z],
\end{equation}
where $\Sigma_{z\varepsilon}\equiv\operatorname{cov}(z,\varepsilon)\in\mathbb{R}^{r\times n}$, $\Sigma_{\varepsilon\varepsilon}=\operatorname{var}(\varepsilon)$, and $\eta$ satisfies $\mathbb{E}[\eta]=0$ and $\mathbb{E}[\eta\varepsilon^\prime]=0$.
From the definition, $u=\boldsymbol{a}_0^\prime \varepsilon$ we have $\Sigma_{zu}=\Sigma_{z\varepsilon}\boldsymbol{a}_0$, where $\Sigma_{zu}\equiv\operatorname{cov}(z,u)\in\mathbb{R}^{r\times r}$. 
We make the following assumption:
\begin{assumption}[Instrument Relevance and Orientation]\label{ass:instrumentrelevant}
The $r\times r$ cross-covariance matrix, 
$$\Phi\equiv  \Sigma_{zu},$$ is non-singular and positive definite.
\end{assumption}
Although positive definiteness implies non-singularity, we state both properties explicitly to emphasize the dual nature of this assumption. The non-singularity of $\Sigma_{zu}$ provides the standard rank condition necessary for instrument relevance. Positive definiteness, on the other hand, secures the correct orientation by requiring a positive association between instruments and their targeted shocks, which is needed for OASIS's objective to maximize these correlations.\footnote{If an instrument, $z_i$ say, is negatively correlated with $u_i$, then substituting $-z_i$ for $z_i$ will make the association positive.} Furthermore, this definiteness condition places a natural bound on the degree of instrument cross-contamination. Mathematically, it requires the diagonal elements (the own-shock signals) to be sufficiently large relative to the off-diagonal elements, ensuring that each instrument $z_i$ is primarily driven by its targeted structural shock $u_i$ rather than being overwhelmed by leakage from other shocks.

We have introduced a special notation for $\Phi = \Sigma_{zu}$ because it plays a central role, and we will refer to $\Phi$ as the \emph{signal matrix} because it characterizes the strength of the instrumental variables. 

Recall our definition of $u=\boldsymbol{a}_0^\prime\varepsilon$ and $v=\boldsymbol{b}_0^\prime\varepsilon$, where the latter is a vector of auxiliary shocks that satisfies $\operatorname{cov}(u,v)=0$.
\begin{assumption}[Exclusion Restriction]\label{ass:exclusion}
$\operatorname{cov}(z,v)=0$.  
\end{assumption}
We have the following representation.
\begin{lemma}[Linear Projection Representation]\label{lem:LinProject}Suppose that Assumptions \ref{ass:instrumentrelevant} and \ref{ass:exclusion} hold. Then $\Sigma_{z\varepsilon} = \Sigma_{zu}\boldsymbol{a}_0^\prime\Sigma_{\varepsilon\varepsilon}$ and the $\eta$ in the linear projection,
$$z = \mu_z + \Phi u + \eta,$$   
satisfies $\mathbb{E}[\eta]=0$ and $\mathbb{E}[\varepsilon \eta^\prime]=0$. \end{lemma}
From this representation it is evident that correlations between the elements of $z$ can arise in two ways. First, if the elements of $\eta$ are correlated, and second, if $\Phi$ is a non-diagonal matrix. We refer to the latter as \emph{leakage} because $\Phi_{ij}\neq 0$, $i\neq j$, means that $z_i$ is not only correlated with the structural shock, $u_i$, but also ``contaminated'' by $u_j$.

The following Theorem characterizes the condition that ensures that the optimal $\boldsymbol{a}_\ast$ (OASIS) recovers the true structural rotation, $\boldsymbol{a}_0$.

\begin{theorem}\label{thm:OASISrecoversRotation}
Suppose  that Assumptions \ref{ass:instrumentrelevant} and \ref{ass:exclusion} hold and let $\Lambda_{\sigma_z}\equiv\operatorname{diag}(\sigma_{z_1},\ldots,\sigma_{z_r})$. If $M=\Phi^\prime\Lambda_{\sigma_z}^{-1}\Lambda_w$ is symmetric and positive definite, then OASIS perfectly recovers the true structural rotation $\boldsymbol{a}_\ast = \boldsymbol{a}_0$, and we have $M = V \Lambda_\xi V^\prime$, where $\Lambda_\xi$ and $V$ are identical to those in the singular value decomposition $\Xi = U \Lambda_\xi V^\prime$.     
\end{theorem}
Notice that the symmetry condition in Theorem \ref{thm:OASISrecoversRotation} is easily satisfied under standard empirical configurations. Specifically, $M$ is symmetric if either $\Phi$ is strictly diagonal, or if $\Phi$ itself is symmetric and the weighting matrix is strictly proportional to the standard deviations of the instruments, $\Lambda_w \propto \Lambda_{\sigma_z}$ (i.e., $w_i = c \sigma_{z_i}$ for some constant $c > 0$ and all $i=1,\ldots,r$). 

A particularly useful scale-invariant case arises when the proxy variables are pre-standardized such that $\Lambda_{\sigma_z} = I_r$. In this scenario, the symmetry of $\Phi$ combined with equal weighting ($\Lambda_w \propto I_r$) provides a sufficient condition for the symmetry of $M$.

The symmetry condition establishes a direct geometric mapping between the data and the structural parameters, explaining why the standard SVD of $\Xi$ seamlessly recovers the spectral properties of the underlying structural matrix $M$ without requiring any additional non-linear optimization.

\subsection{Optimal Combination of Multiple Instruments}

Next, we turn to the more general situation where we may have multiple instrumental variables to identify each of the structural shocks. Having multiple proxies per shock offers two advantages. First, it enables better signal extraction, linear combination across a block of correlated proxies can effectively reduce the idiosyncratic noise in individual proxies, thus generating a ``stronger'' composite instrument that maximizes the signal-to-noise ratio. Second, multiplicity introduces testable implications for instrument validity, which is conceptually analogous to the overidentifying restrictions test ($J$-test) in the classical Two-Stage Least Squares (2SLS) framework. 

As before we define $\eta$ from the linear projection, $Z=\mu_Z+\Sigma_{Z\varepsilon} \Sigma_{\varepsilon\varepsilon}^{-1} \varepsilon + \eta$, such that $\mathbb{E}[\eta] = 0$ and $\mathbb{E}[\eta\varepsilon]=0$. The exclusion restriction now reads:
\newtheorem*{assumption2prime}{Assumption 2'}
\begin{assumption2prime}[Exclusion Restriction]
$\operatorname{cov}(Z,v)=0$.  
\end{assumption2prime}

\begin{lemma}\label{lem:ZeInv(ee)eZ}
Let Assumption 2' hold. Then the vector of proxies admits the representation
$$
Z=\mu_Z+\boldsymbol{\Phi}u+\eta,
\qquad
\boldsymbol{\Phi}\equiv \operatorname{cov}(Z,u),
$$
where $\mathbb{E}[\eta]=0$ and $\mathbb{E}[\eta\varepsilon^\prime]=0$.
Moreover, $\Sigma_{Z\varepsilon}=\boldsymbol{\Phi} \boldsymbol{a}_0^\prime \Sigma_{\varepsilon\varepsilon}$ and $\Sigma_{Z\varepsilon}\Sigma_{\varepsilon\varepsilon}^{-1}\Sigma_{\varepsilon Z}
        =\boldsymbol{\Phi}\boldsymbol{\Phi}^\prime$.
\end{lemma}

Let $k_i=\operatorname{dim}(Z_i)$ be the number of instrumental variables intended to identify the $i$-th structural shock $u_i$, such that $Z$ has dimension $k=\sum_{i=1}^r k_i$. We partition $Z$ and $\boldsymbol{\Phi}$ accordingly,
$$
Z = 
\left(
\begin{array}{c}
    Z_1\\
    \vdots\\
    Z_r
\end{array}
\right),\quad
\boldsymbol{\Phi} = \left(
\begin{array}{ccc}
  \boldsymbol{\phi}_{11}   &\cdots & \boldsymbol{\phi}_{1r}  \\
   \vdots  & &\vdots\\
  \boldsymbol{\phi}_{r1}   &\cdots & \boldsymbol{\phi}_{rr}  \\
\end{array}\right)
\in \mathbb{R}^{k \times r}
$$

Our objective is to construct an $r$-dimensional composite instrument vector. We assume that the $i$-th block is primarily associated with the $i$-th structural shock, $u_i$, 
such that the composite instruments take the form:
$$
\bar{z} = (\bar{z}_1, \bar{z}_2, \ldots, \bar{z}_r)^\prime \in \mathbb{R}^{r}, \quad \text{where} \quad \bar{z}_i = \beta_i^\prime Z_i \in \mathbb{R},\text{ for }i=1,\ldots,r.
$$
To identify the (sign of) $\beta_i^\ast$, we assume that the first element of $Z_i$ is known to be positively correlated with the target shock. We refer to this proxy variable as an \emph{anchor proxy}. 
\newtheorem*{assumption1prime}{Assumption 1'}
\begin{assumption1prime}[Anchor Proxy]For each block $i=1,\ldots,r$, 
    $\boldsymbol{\phi}_{ii,1} > 0$ and $\Sigma_{Z_i Z_i}$ is nonsingular.
\end{assumption1prime}
Once $\beta_i^\ast$, $i=1,\ldots,r$ have been fully determined, we arrive at the condensed system:
$$
    \bar{z} = \mu_{\bar{z}}+\bar{\Phi} u + \bar{\eta}, \quad \text{where} \quad  \bar{\Phi} = B^\prime \boldsymbol{\Phi} , \quad B=\operatorname{diag}(\beta_1,\ldots, \beta_r)
$$
where the problem has the structure analyzed above. 
$$
    \boldsymbol{a}_* = \arg \max_{\boldsymbol{a} \in \mathcal{A}_r} \ g(\boldsymbol{a})=\sum_{i=1}^r w_i \operatorname{corr}(u_i,\bar{z}_i).
$$
However, an important difference is that $\operatorname{corr}(u_i,\bar{z}_i)$ depends on both $\beta_i$ and $\boldsymbol{a}$. We will establish conditions that ensure these can be determined and that OASIS  recovers the structural rotation, $\boldsymbol{a}_\ast=\boldsymbol{a}_0$.

Without loss of generality, we restrict attention to linear combinations that generate composite instruments with unit variance, i.e. $\beta_i^\prime\Sigma_{Z_i Z_i}\beta_i=1$, and we define the optimal $\beta_i$ by
\begin{equation}\label{eq:MaxCorrObjective}
\beta_i^* =\underset{\beta_i^{\prime} \Sigma_{Z_i Z_i} \beta_i=1}{\operatorname{argmax}}\operatorname{corr}(u_i,\beta_i^\prime Z_i),\qquad\text{for }i=1,\ldots,r. 
\end{equation}
This is obviously complicated by the latent nature of $u_i$.
However, the squared correlation can be expressed as $\operatorname{corr}^2(u_i,\beta_i^\prime Z_i)={\beta_i^{\prime} (\boldsymbol{\phi}_{ii}\boldsymbol{\phi}_{ii}^\prime)\beta_i}$, where we used $\beta_i^{\prime} \Sigma_{Z_i Z_i} \beta_i=1$ and $\operatorname{cov}(Z_i,u_i)=\boldsymbol{\phi}_{ii}$. 
So, a major step towards solving (\ref{eq:MaxCorrObjective}) is to maximize the Rayleigh coefficient: 
\begin{equation}\label{eq:Rayleigh}
    \max_b\frac{b^{\prime} (\boldsymbol{\phi}_{ii}\boldsymbol{\phi}_{ii}^\prime)b}{b^\prime \Sigma_{Z_i Z_i} b}.
\end{equation}
The remaining hurdles to this problem are that $\boldsymbol{\phi}_{ii}$ is unobserved and that the solution to (\ref{eq:Rayleigh}) is not unique. We address these issues next.

\subsubsection{Baseline Case for Multiple Instruments}
The vector of proxy variables, $Z_i$, should primarily be tied to the target shock, $u_i$, but may be functionally related to the entire structural vector, which we refer to as \emph{leakage}. First, we consider a simplified baseline setup without leakage, such that $Z_i\in\mathbb{R}^{k_i}$ load exclusively on the targeted shock, $u_i$, $i=1,\ldots,r$. 
\begin{assumption}[No leakage]\label{ass:NoLeakage}
    The cross-loadings are strictly zero: $\boldsymbol{\phi}_{ij} = \mathbf{0}$ for $i\neq j$.
\end{assumption}
This assumption implies that $\operatorname{cov}(Z_i,u_j)=0$ for $i\neq j$, which is identical to \citet[eq.~6.ii]{StockWatson:2012}. 

Analogous to Lemma \ref{lem:LinProject}, we have the following results: 
\begin{lemma}\label{lem:Baseline}
    Let Assumptions 1', 2', and \ref{ass:NoLeakage} hold. Then
    $\Sigma_{\varepsilon Z_i}  = \Sigma_{\varepsilon \varepsilon} \boldsymbol{a}_{0, i} \boldsymbol{\phi}_{ii}^{\prime}$ 
    and 
    $$
        Z_i = \mu_{Z_i}+\boldsymbol{\phi}_{ii} u_i + \eta_i, \quad \text{where} \quad \boldsymbol{\phi}_{ii} \in \mathbb{R}^{k_i},
    $$
where $\mathbb{E}[\varepsilon\eta_i^\prime]=\mathbf{0}$ for $i=1,\ldots,r$.
\end{lemma}

The first identity in Lemma \ref{lem:Baseline} resolves the unobservable nature of $\boldsymbol{\phi}_{ii}$, because the objective in (\ref{eq:Rayleigh}) simplifies to the observable Rayleigh quotient:
$$
f_i(b)\equiv \frac{b^{\prime} (\boldsymbol{\phi}_{ii}\boldsymbol{\phi}_{ii}^\prime)b}{b^\prime \Sigma_{Z_i Z_i} b}
=
\frac{b^{\prime} \left(\Sigma_{Z_i \varepsilon} \Sigma_{\varepsilon \varepsilon}^{-1} \Sigma_{\varepsilon Z_i}\right) b}{b^{\prime} \Sigma_{Z_i Z_i} b},
$$
where $H_i\equiv\Sigma_{Z_i \varepsilon} \Sigma_{\varepsilon \varepsilon}^{-1} \Sigma_{\varepsilon Z_i}$ and $\Sigma_{Z_i Z_i}$ can be estimated from the data. 

This leads to a generalized eigenvalue problem, which is common in the related econometrics literature.\footnote{Examples include: Canonical Correlation Analysis \citep{Hotelling:1936}, the Limited Information Maximum Likelihood (LIML) estimator \citep{AndersonRubin:1949}, and maximum likelihood cointegration analysis \citep{Johansen:1988}.} 

Let $q_i^\ast$ denote the principal eigenvector of 
    $$
        G_i\equiv \Sigma_{Z_i Z_i}^{-1/2}H_i \Sigma_{Z_i Z_i}^{-1/2},
    $$
    and let $\lambda_i^\ast=\lambda_{\max}(G_i)$ denote the corresponding eigenvalue, $i=1,\ldots,r.$ 
    
    Under Assumption 3, the largest eigenvalue satisfies
\begin{equation}\label{eq:lambdaAlphaSq}
    \lambda_i^\ast=\alpha_i^2 \equiv \boldsymbol{\phi}_{ii}^\prime \Sigma_{Z_iZ_i}^{-1} \boldsymbol{\phi}_{ii},
\end{equation}
such that the largest eigenvalue can be interpreted as the R-squared from regressing $u_i$ on the vector of proxy variables $Z_i$ and a constant.
Next, we define 
    $$
        \tau_i \equiv \operatorname{sign}([\Sigma_{Z_i Z_i}^{1/2} q_i^\ast]_1)\in\{-1,1\}.
    $$
\begin{theorem}\label{thm:OptimalBeta}
    Let Assumptions 1', 2', and 3 hold. 
    Then 
    \begin{equation}\label{eq:betaStar}
        \beta_i^* 
        = \tau_i \Sigma_{Z_i Z_i}^{-1/2} q_i^\ast,
    \end{equation}
    is the optimal solution to (\ref{eq:Rayleigh}), which satisfies $\operatorname{var}(\bar{z}_i)=1$ and is consistent with $[\boldsymbol{\phi}_{ii}]_1>0$, $i=1,\ldots,r$. 

    Moreover, the composite instruments are such that 
    $
        \bar{\Phi} = \operatorname{cov}(\bar{z},u)
    $
    is a diagonal matrix with strictly positive elements, such that OASIS applied with $\bar{z}$ recovers the structural rotation: $\boldsymbol{a}_\ast = \boldsymbol{a}_0$.  
\end{theorem}

\subsubsection{Proportional Leakage Case for Multiple Instruments}

The baseline combination strategy above assumes that the instruments in block $Z_i$ load exclusively on the targeted shock $u_i$. This assumption can fail in empirical settings because proxies may be influenced by a wider range of structural shocks, such that: 
\begin{equation}\label{eq:Zleakage}
    Z_i = \mu_{Z_i}+\sum_{j=1}^r \boldsymbol{\phi}_{ij} u_j + \eta_i\in \mathbb{R}^{k_i}.
\end{equation}
We refer to the case $\boldsymbol{\phi}_{ij}\neq 0$ for $i\neq j$ as leakage, and we consider the case with symmetric leakage. As before we assume that own-shock loading vectors are strictly positive.
\newtheorem*{assumption1primeprime}{Assumption 1''}
\begin{assumption1primeprime}[Proportional Leakage]
$\boldsymbol{\phi}_{ii,1} > 0$ and the cross-loadings satisfy $\boldsymbol{\phi}_{ij} = s_{ij}\boldsymbol{\phi}_{ii}$ for some scalar parameter $s_{ij} \in \mathbb{R}$.
\end{assumption1primeprime}

\begin{lemma}\label{lem:ZieInv(ee)eZi}Given Assumptions 1'' and 2'. We have
    $$
    \Sigma_{Z_i\varepsilon}\Sigma_{\varepsilon\varepsilon}^{-1}\Sigma_{\varepsilon Z_i}
    = s_{i\bullet}^2 \boldsymbol{\phi}_{ii}\boldsymbol{\phi}_{ii}^\prime,
    $$
    where $s_{i\bullet}^2 \equiv\sum_{j=1}^r s_{ij}^2 \geq 1$.
\end{lemma}
Lemma \ref{lem:ZieInv(ee)eZi} shows that the Rayleigh ratio for this problem is proportional to that in the baseline case, such that $\beta_i^\ast$ is not affected by leakage and is therefore identical to the solution in (\ref{eq:betaStar}). Consequently, the composite instruments, $\bar{z}=B^{\ast\prime} Z$, are identical to those in the baseline case, and we have $\bar{\Phi}=\operatorname{cov}(\bar{z},u)$.

In the presence of leakage, $\lambda_i^\ast=\lambda_{\max}(G_i)$ need not be equal to $\alpha_i^2=\boldsymbol{\phi}_{ii}^\prime \Sigma_{Z_i Z_i}^{-1} \boldsymbol{\phi}_{ii}$. Instead we have  
\begin{equation}\label{eq:lambdaAlphaSqLeakage}
    \lambda_i^\ast=s_{i\bullet}^2\alpha_i^2\geq \alpha_i^2.
\end{equation}
which is an implication of Lemma \ref{lem:ZieInv(ee)eZi}, and we make the following assumptions about the structural leakage coefficients. 
\newtheorem*{assumption3prime}{Assumption 3'}
\begin{assumption3prime}[Structural Independence]
The structural leakage matrix, 
$$
    S =\left[\begin{array}{cccc}1 & s_{12} & \ldots & s_{1 r} \\ s_{21} & 1 & \ldots & s_{2 r} \\ \vdots & \vdots & \ddots & \vdots \\ s_{r1} & s_{r2} & \ldots & 1\end{array}\right]\in \mathbb{R}^{r \times r},
$$
is symmetric and positive definite.    
\end{assumption3prime}

\begin{theorem}[Structural Identification with Proportional Leakage]\label{thm:ProxyOasisLeakage}
    Let Assumptions 1'', 2', and 3' hold, and let $\beta_i^\ast$ be as in (\ref{eq:betaStar}). 
    Then $$\operatorname{cov}(\bar{z},u)=\bar{\Phi} = \Lambda_\alpha S,$$
    where $\Lambda_\alpha=\operatorname{diag}(\alpha_1,\ldots,\alpha_r)$. 
    
    Moreover, the diagonal signal matrix $\Lambda_\alpha$ is uniquely identified from the observable composite matrix $H \equiv \Sigma_{\bar{z}\varepsilon}\Sigma_{\varepsilon\varepsilon}^{-1}\Sigma_{\varepsilon\bar{z}}$, as the unique solution to 
    \begin{equation}\label{eq:Lambdaalpha}
        \operatorname{diag}( [\Lambda_\alpha^{-1} H \Lambda_\alpha^{-1}]^{1/2} ) = \iota,
    \end{equation}
    where $\iota$ is the $r$-dimensional vector of ones. 

    If we set $\Lambda_w=\Lambda_\alpha^{-1}$, then OASIS recovers the structural rotation, $\boldsymbol{a}_\ast = \boldsymbol{a}_0$ and $S = V \Lambda_\xi V^\prime$, where $\Lambda_\xi$ and $V$ are identical to those in the singular value decomposition $\bar{\Xi} = C_{\varepsilon \varepsilon}^{-1/2}C_{\varepsilon \bar{z}}\Lambda_w=U \Lambda_\xi V^\prime$.    
\end{theorem}
Equation (\ref{eq:Lambdaalpha}) identifies $\Lambda_\alpha$ which in turn identifies $S=[\Lambda_\alpha^{-1} H \Lambda_\alpha^{-1}]^{1/2}$. The following algorithm can be used to compute $\Lambda_\alpha$. 
\begin{remark}Initialize $\alpha_i^{0}$ to a positive value (e.g. 1) and apply the recursion 
$$
\log \alpha_i^{(k+1)} = \log \alpha_i^{(k)} + \log \left[\left(\Lambda_{\alpha^{(k)}}^{-1} H \Lambda_{\alpha^{(k)}}^{-1}\right)^{1 / 2}\right]_{i i},
$$
for $k=1,2,\ldots$, which rapidly converges to the unique scaling matrix that standardizes the diagonal of the structural impact matrix.
\end{remark}
The algorithm is structurally identical to the fixed-point iteration used to compute the Generalized Fisher Transformation (GFT) of correlation matrices in \citet{ArchakovHansen:Correlation}. In the GFT framework, an analogous recursion is utilized to find the unique diagonal elements that force the matrix exponential to yield a strict unit diagonal. 

The ability to empirically recover $S$ from observable data is not merely a theoretical curiosity; it resolves a pervasive issue in applied macroeconomic research. To preview the empirical relevance of Theorem 6, our applications demonstrate severe cross-contamination among macroeconomic proxy variables. For instance, in Section 5.2 we apply our framework to \citet{StockWatson:2012} and uncover leakage effects as large as $0.661$ between economic uncertainty and financial risk. Furthermore, in an application to \citet{MertensRavn:2013}, we estimate a cross-shock leakage parameter of roughly $0.234$ between their two narrative tax instruments. This pervasive non-orthogonality directly invalidates classical exclusion assumptions and highlights the practical necessity of recovering $S$ for robust structural identification.

More generally, the structural leakage matrix $S$ in the Proxy-VAR framework plays a conceptual role analogous to the structural impact matrix in standard SVARs. While Assumption 3$^\prime$ imposes symmetry on $S$ to achieve exact point identification, the OASIS framework is sufficiently flexible to accommodate alternative structural restrictions. Other theoretically motivated structure on $S$ can be entertained, provided there exists a diagonal weighting matrix $\Lambda_{w}$ such that the composite signal matrix $\Lambda_{w}\Lambda_{\alpha}S$ is symmetric and positive definite.

\subsection{Rank Test for Overidentification}

Under the exclusion restriction, the assumption of proportional leakage (Assumption 1$^{\prime\prime}$) imposes a strict rank-one structure on the observable cross-covariance matrix. This joint restriction yields a testable implication conceptually analogous to the overidentifying restrictions test ($J$-test) in the classical Two-Stage Least Squares (2SLS) framework, which jointly tests instrument validity and exogeneity.

Specifically, Lemma \ref{lem:ZieInv(ee)eZi} establishes that the cross-covariance matrix $H_i=\Sigma_{Z_i\varepsilon}\Sigma_{\varepsilon\varepsilon}^{-1}\Sigma_{\varepsilon Z_i} = s_{i\bullet}^2 \boldsymbol{\phi}_{ii}\boldsymbol{\phi}_{ii}^\prime$ has a rank of exactly one. Consequently, our observable matrix $G_i=\Sigma_{Z_i Z_i}^{-1/2}H_i\Sigma_{Z_i Z_i}^{-1/2}$ also has a rank of one in the population, which the only non-trivial eigenvalue being $\lambda_1=s_{i \bullet}^2 \alpha_i^2$, while the remaining $k_i - 1$ eigenvalues are exactly zero.

We can formally test whether the $k_i - 1$ smallest eigenvalues are jointly zero using the standard trace statistic from canonical correlation analysis \citep{Anderson:1951}:
\begin{equation}\label{eq:J_trace}
    \mathcal{J}_i = -T \sum_{m=2}^{k_i} \log(1 - \hat{\lambda}_m),
\end{equation}
where $T$ is the sample size and $\hat{\lambda}_1 > \hat{\lambda}_2 \ge \dots \ge \hat{\lambda}_{k_i}$ denote the sorted sample eigenvalues of the estimated matrix $\hat{G}_i$. 
Under the null hypothesis of proportional leakage and conditional homoskedasticity, $\mathcal{J}_i$ follows an asymptotic $\chi^2$ distribution with $(k_i - 1)(n - 1)$ degrees of freedom, where $n$ is the dimension of the VAR system. 

If the test strongly rejects the null, it provides statistical evidence that the proxy variables within the block $Z_i$ capture fundamentally different structural dynamics. In such cases, collapsing them into a single composite instrument via $\beta_i^\ast$ will result in a loss of structural identification.

Anderson's rank-test is not robust to the heteroskedasticity and autocorrelation that is prevalent in economic time series. We will therefore adopt a robust variant by \citet{KleibergenPaap:2006}. Their rank-test is applied to $\Pi_i = \mathbb{E}[Z_{i,t}\varepsilon_t^\prime]$, using the fact that 
$\operatorname{rank}G_i=\operatorname{rank}\Pi_i$ under the assumption that $\Sigma_{\varepsilon\varepsilon}$ and $\Sigma_{Z_i Z_i}$ have full rank. This facilitates the GMM framework with straightforward computation of robust standard errors.

Our implementation of the Kleibergen-Paap test is as follows: We obtain the SVD of the sample cross-covariance matrix 
$
    \hat{\Pi}_i = \frac{1}{T} \sum_{t=1}^T Z_{i,t} \hat{\varepsilon}_t^\prime\in\mathbb{R}^{k_i\times n}
$
$$
    \hat{\Pi}_i = U \Lambda_\pi V^\prime = \pi_1 U_1  V_1^\prime+U_2 \tilde{\Lambda}_\pi V_2^\prime,
$$ where the singular values in the diagonal of $\Lambda_\pi$ are sorted in descending order and we partially suppress the dependence on $i$. Here $\tilde{\Lambda}_\pi$ contains the  $k_i - 1$ smallest singular values, and $U_2\in\mathbb{R}^{k_i \times k_i - 1}$ and $V_2\in\mathbb{R}^{n \times (k_i - 1)}$ are the corresponding submatrices of $U$ and $V$, respectively. 

The empirical moment conditions vector is then constructed by projecting the sample covariance matrix into this restricted noise space:
$$    
    \hat{m}_i = \operatorname{vec}(U_2^\prime \hat{\Pi}_i V_2)
$$
Next, define $\zeta_{i,t} \equiv \varepsilon_t \otimes Z_{i,t}$, and let $\hat\Sigma_{\zeta_i }$ denote a consistent estimate of the long-run variance of $\{\zeta_{it}\},$\footnote{In our empirical section, we estimate the long-run variance using a Parzen kernel with the automatic bandwidth selection procedure of \cite{Andrews:1991}.} and define $\hat{\Omega}_i=(V_2 \otimes U_2)^\prime \hat{\Sigma}_{\zeta_i} (V_2 \otimes U_2)$ and let $\hat{\Omega}_i^{\#}$ denote its generalized inverse.

The test statistic is now given by, 
$$
    J_{\mathrm{KP},i} = T \cdot \hat{m}_i^\prime \hat{\Omega}_i^{\#} \hat{m}_i,
$$
which is asymptotically distributed as a $\chi^2_{(k_i - 1)(n - 1)}$ for large $T$, see \citet{KleibergenPaap:2006} for details.

\section{Revisiting Empirical SVAR Studies}\label{sec:Studies}

We conduct an extensive review of existing studies that employ a Cholesky decomposition in VARs to identify structural shocks. The studies are listed in  Table \ref{tab:shock_studies}, organized by topic (Monetary, Fiscal, Uncertainty, Financial, Oil, and Sectoral). We include a brief summary of these articles that predominantly use impact timing restrictions to reveal impulse responses.
\begin{table}[htbp!]
    \caption{Selected Structural VAR Studies with Causal-Ordering Restrictions by Topic}
    \label{tab:shock_studies}
    {\setstretch{1.00}\footnotesize
        \begin{tabularx}{\textwidth}{@{}p{1.4cm}X @{}}
            \toprule
            \multicolumn{2}{@{}l}{\textbf{Topic/[Abbreviation]/Reference/Notes}} \\[1mm]
            \midrule
            \multicolumn{2}{c}{\textbf{Monetary Shocks}}\\
            {[B86]} &\citet{Bernanke:1986}. VAR using monetary aggregates and interest rates as policy indicators. \\
            {[S95]} &\citet{strongin1995}. Identifies monetary shocks via Fed's component of nonborrowed reserves. \\
            {[LSZ96]} &\citet{Leeperetal1996}.  SVAR analyzing monetary policy; emphasizes sensitivity to identification assumptions and robustness to ordering. \\
            {[CEE99]} &\citet{ChristianoEichenbaumEvans:1999}. SVAR using federal funds rate as monetary policy instrument. \\
            {[CEE05]} &\citet{ChristianoEichenbaumEvans:2005}.   SVAR using interest rate shocks and impulse response matching to validate DSGE models. \\
            {[BL09]}&\citet{bjornland2009}.  SVAR identifying monetary shocks via interest rates and stock prices. \\           
            \multicolumn{2}{c}{\textbf{Fiscal Shocks}}\\
            {[BP02]}&\citet{blanchard2002}. SVAR using government spending and net taxes with timing restrictions. \\
            {[RZ11]} &\citet{RossiZubairy:2011}.  SVAR using government spending, tax revenue, and interest rates. \\
            {[FG16]}&\citet{ForniGambetti:2016}. Uses forecast revisions (SPF) and government spending as shock proxy. \\           
            \multicolumn{2}{c}{\textbf{Uncertainty Shocks}}\\
            {[B09]}& \citet{bloom2009}.  SVAR using stock market volatility (VXO) as proxy for uncertainty shocks. \\
            {[CCG14]}&\citet{caggiano2014}.  VAR using VIX and forecast dispersion to measure uncertainty. \\
            {[BB17]}& \citet{basu2017}.  SVAR identified via stock market volatility (VXO). \\
            {[BO23]}& \citet{bonciani2023uncertainty}. VAR with uncertainty shocks proxied by macro uncertainty measures estimated by \cite{Juradoetal2015}; compared to DSGE. \\[5mm]            
            \multicolumn{2}{c}{\textbf{Financial Shocks}}\\
            {[GZ12]}& \citet{gilchrist2012}.  VAR using credit spreads and bond premia as financial shock indicators. \\
            {[FGMS24]}&\citet{ForniGambettiMaffei-FaccioliSala:2024}. VAR with nonlinear financial shock identification using a Vector Moving Average (VMA) model. \\           
            \multicolumn{2}{c}{\textbf{Oil Price Shocks}}\\
            {[LS04]}& \citet{leduc2004}.  SVAR using real oil price as shock variable.\\
            {[LP18]}&\citet{lorusso2018causes}.  SVAR using global oil prices to assess UK macro responses. \\[5mm]            
            \multicolumn{2}{c}{\textbf{Sectoral}}\\
            {[FGKV25]}& \citet{Fry-McKibbinGreenwood-NimmoKimVolkov:2025}. SVAR for Australia, controlling for import penetration from China, in the sectoral output context. \\
            \bottomrule
        \end{tabularx}
        \vspace{0.3mm}\\        
        \noindent\textit{Notes:} Empirical studies that employed VARs or SVARs to identify and trace the effects of macroeconomic shocks. Aside from the first two studies, the results were reported to be robust to Cholesky reordering.
    }
\end{table}

\subsection{OASIS and Cholesky Results}

Table \ref{tab:EmpiricalResults} provides detailed information about each of the empirical studies. We report the values of $\bar\rho_\ast$ and $\bar\rho_\mathrm{c}$ for each of the studies, where we have used the same ordering of variables as in the original articles. We also report the range of $\bar\rho_\mathrm{c}$ over all Cholesky orderings and descriptive statistics, such as the dimension of the VAR, $n$, the correlation between the two types of structural shocks ($u^\ast$ and $u$), $\bar\rho_{\ast,\mathrm{c}}$, the average absolute correlation between reduced-form shocks, $\|C\|_1^{\star}=\tfrac{1}{n(n-1)}\sum_{i\neq j}|C_{ij}|$, the ratio, $\tfrac{1-\bar{\rho}_\mathrm{c}}{1-\bar{\rho}_\ast}$, (which should be about two according to Theorem~\ref{thm:Proximity}), and $d(C)=\tfrac{1}{n}\sum_{i\neq j}C_{ij}^2$.
\begin{table}[htbp!]
    \caption{OASIS and Cholesky Identification in SVARs}\label{tab:EmpiricalResults}
    {\setstretch{1.00}\footnotesize
        \begin{tabularx}{\textwidth}{@{}>{\hsize=1.6\hsize}X
        >{\hsize=0.15\hsize}X 
        >{\hsize=1.75\hsize\centering\arraybackslash}X
        >{\hsize=0.75\hsize\centering\arraybackslash}X
        >{\hsize=0.75\hsize\centering\arraybackslash}X
        >{\hsize=0.75\hsize\centering\arraybackslash}X
        >{\hsize=1.75\hsize\centering\arraybackslash}X
        >{\hsize=0.75\hsize\centering\arraybackslash}X
        >{\hsize=1.00\hsize\centering\arraybackslash}X
        >{\hsize=0.75\hsize\centering\arraybackslash}X@{}}
        \toprule
         &  & \multicolumn{1}{c}{OASIS} & \multicolumn{3}{c}{Cholesky} &  &  & &  \\
        Study & 
        $n$ & \multicolumn{1}{c}{$\bar{\rho}_\ast$} & \multicolumn{1}{c}{$\bar{\rho}_\mathrm{c}$} & min & max & $\bar{\rho}_{\ast,\mathrm{c}}$  & $\|C\|_1^{\star}$ & $\tfrac{1-\bar{\rho}_\mathrm{c}}{1-\bar{\rho}_\ast}$ & $d(C)$\\[3mm]
        \midrule
             \\[-2mm]
            {[B86]} & 6 & 0.965 & 0.935 & 0.933 & 0.936     & 0.970 & 0.17 & 1.88& 0.29  \\
            \ \ --- & 6 & 0.962 & 0.930 & 0.928 & 0.931       & 0.968 & 0.18 & 1.87& 0.32  \\
            {[S95]} & 5 & 0.988 & 0.977 & 0.976 & 0.977     & 0.988 & 0.13 & 1.99& 0.09  \\
            {[LSZ96]} & 4 & 0.987 & 0.975 & 0.975 & 0.975   & 0.987 & 0.14 & 1.97& 0.10  \\
            {[CEE99]} & 7 & 0.967 & 0.933 & 0.932 & 0.936   & 0.963 & 0.13 & 2.02& 0.24  \\
            {[CEE05]} & 9 & 0.925 & 0.860 & 0.857 & 0.869    & 0.935 & 0.24 & 1.88& 0.70  \\
            {[BL09]} & 5 & 0.995 & 0.991 & 0.991 & 0.991    & 0.996 & 0.09 & 1.95& 0.04  \\[2mm]
            {[BP02]} & 3 & 0.982 & 0.966 & 0.965 & 0.966    & 0.983 & 0.23 & 1.95& 0.14  \\
            \ \ --- & 7 & 0.903 & 0.785 & 0.757 & 0.827      & 0.831 & 0.26 & 2.22& 0.63  \\
             {[RZ11]} & 8 & 0.915 & 0.834 & 0.815 & 0.849   & 0.908 & 0.22 & 1.95& 0.61  \\
             {[FG16]} & 7 & 0.969 & 0.935 & 0.934 & 0.939   & 0.963 & 0.15 & 2.09& 0.22  \\
            \ \ --- & 8 & 0.968 & 0.935 & 0.934 & 0.939     & 0.965 & 0.14 & 2.04& 0.24  \\[2mm]
             {[B09]} & 8 & 0.965 & 0.936 & 0.932 & 0.936    & 0.970 & 0.12 & 1.83& 0.28  \\
             {[CCG14]} & 4 & 0.988 & 0.976 & 0.975 & 0.976  & 0.988 & 0.13 & 1.98& 0.09  \\
             {[BB17]} & 8 & 0.935 & 0.871 & 0.869 & 0.881   & 0.933 & 0.22 & 1.99& 0.53  \\
             {[BO23]} & 9 & 0.945 & 0.897 & 0.896 & 0.904   & 0.952 & 0.22 & 1.88& 0.54  \\[2mm]
             {[GZ12]} & 8 & 0.944 & 0.895 & 0.894 & 0.897   & 0.951 & 0.23 & 1.87& 0.51  \\
             {[FGMS24]} & 6 & 0.976 & 0.954 & 0.954 & 0.955 & 0.978 & 0.13 & 1.92& 0.20  \\[2mm]
             {[LS04]} & 5 & 0.987 & 0.975 & 0.975 & 0.975   & 0.988 & 0.12 & 1.95& 0.10  \\
             {[LP18]} & 3 & 0.999 & 0.998 & 0.998 & 0.998   & 0.999 & 0.06 & 1.99& 0.01  \\
            \ \ --- & 3 & 0.997 & 0.994 & 0.994 & 0.994  & 0.997 & 0.08 & 1.99& 0.02  \\[2mm]
            {[FGKV25]} & 14 & 0.959 & 0.920 & 0.916 & 0.922  & 0.958 & 0.12 & 1.98& 0.32 \\[2mm]
        \bottomrule
        \end{tabularx}
        \vspace{0.3em}\\
        \noindent\textit{Notes:} Abbreviations for each study appear in the first column. $n$ is the dimension of the VAR, $\bar\rho_\ast$ is the average correlation between structural shocks and reduced-form shocks for OASIS. The corresponding correlation for Cholesky is denoted $\bar\rho_\mathrm{c}$ and min and max give the range of correlations for all variable permutations. $\bar{\rho}_{\ast,\mathrm{c}}$ is the average correlation between baseline shocks implied by OASIS and identified Cholesky shocks. $\|C\|_1^{\star}$ is the average absolute correlation in $C$.
    }
\end{table}

Many studies have a correlation matrix that is relatively close to the identity matrix. For instance, those of \citet{bjornland2009} and \citet{lorusso2018causes} have an average absolute correlation of $\|C\|_1^{\ast}<0.10$. This explains that any Cholesky decomposition in these studies will result in a nearly perfect correlation, $\bar\rho>0.99$, between Cholesky-based structural shocks and reduced-form shocks. 
It is interesting that the correlations between the reduced-form shocks tend to be small. This is not a feature of least-squares estimation of vector autoregressions (VARs), because separate equation-by-equation estimation produces identical estimates and residuals. Rather, it suggests that reduced-form shocks in these models tend to be weakly correlated. This is helpful because when the reduced-form shocks are nearly orthogonal, the choice of structural rotation has a less drastic impact on the resulting shock series.
We observe that $d(C)$ is particularly small in the studies of oil price shocks, which could stem from the exogeneity of oil price movements. 

As predicted by Theorem~\ref{thm:Proximity}, studies with large $d(C)$, such as \citet{ChristianoEichenbaumEvans:2005}, \citet{blanchard2002}, and \citet{RossiZubairy:2011}, have the largest differences between OASIS and Cholesky. 

\begin{figure}[htbp!]
\centering{}\includegraphics[width=0.95\textwidth]{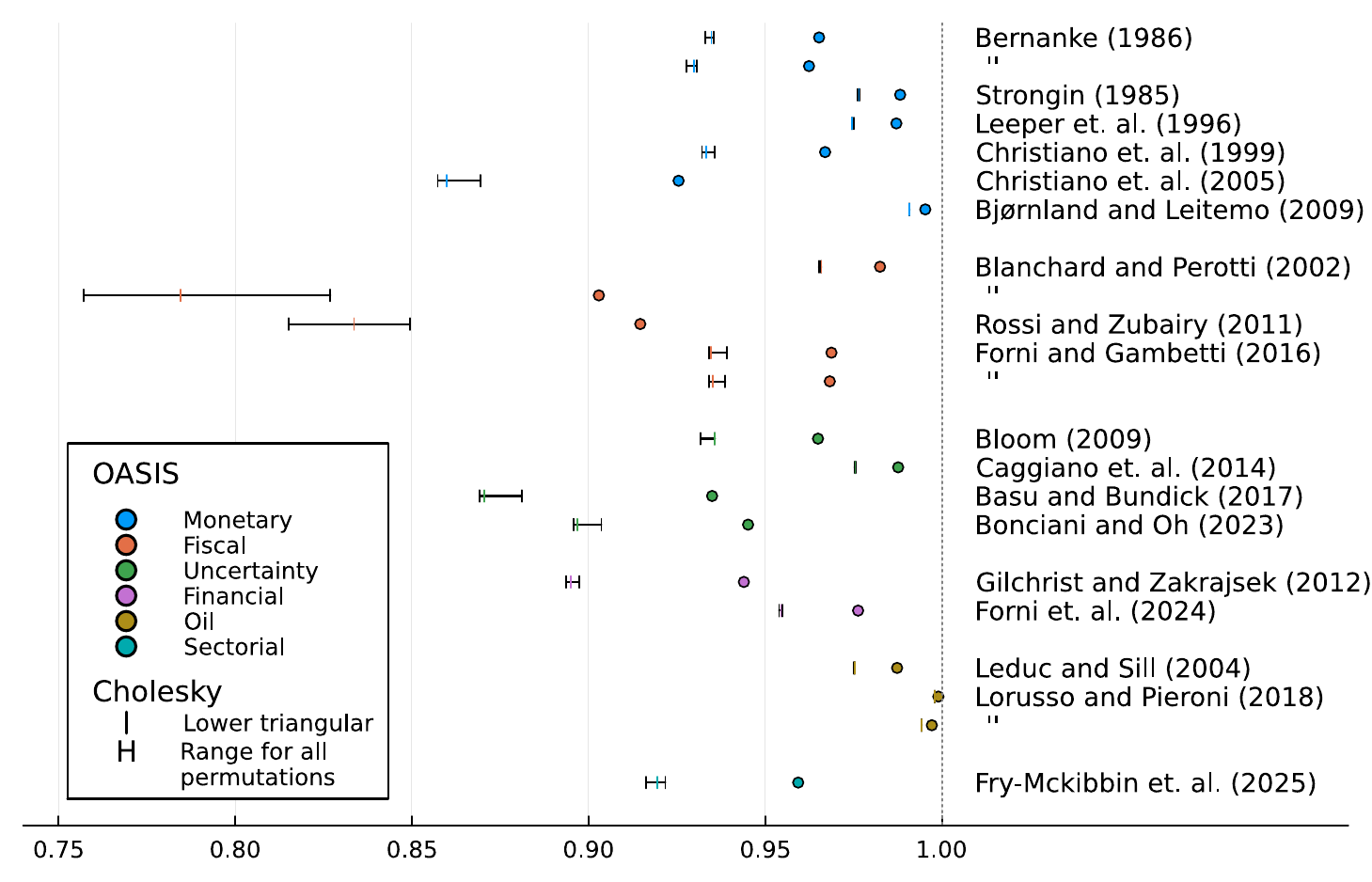}
\caption{{\small{} Selected studies. OASIS is shown for all; some Cholesky values lie below the range shown. Correlations are remarkably high, with a narrow range across Cholesky orderings.\label{fig:AveCorrAll}}}
\end{figure}

Figure \ref{fig:AveCorrAll} is an illustration of some of the results in Table \ref{tab:EmpiricalResults}. We plot the values of $\bar\rho_\ast$ and $\bar\rho_\mathrm{c}$ for each of the studies and use bars to indicate the range of Cholesky outcomes. 
It is striking how large the correlations between structural shocks and reduced-form shocks are across all studies.  All OASIS correlations are above 90\% and 15 of the 22 SVAR specifications have OASIS correlations above 95\%. The same fifteen specifications have Cholesky correlations above 90\%. The range of Cholesky correlations, by considering all possible permutations of the $n$ variables, is also fairly narrow, and for the nine studies where the Cholesky correlation is above 95\%, all permutations have nearly identical average correlations, as predicted by Theorem~\ref{thm:Proximity}. 
Only two studies, the second study with $n=7$ in \citet{blanchard2002} and \citet{RossiZubairy:2011}, have somewhat modest values of $\bar{\rho}_\mathrm{c}$, and these are also the only two studies where the choice of Cholesky ordering can affect the average correlation to some extent. This does not reflect a weakness of these studies but is simply a consequence of the covariance structure of the variables under investigation. 
Even the largest VAR with $n=14$ has a relatively narrow range, from 91.63\% to 92.17\%, for the average correlation between structural shocks and reduced-form shocks. This is remarkable because there are more than 87 billion different Cholesky orderings to consider, yet all of them have nearly the same value of $\bar\rho_\mathrm{c}$.

Despite the average correlation being very similar for all Cholesky orderings, the resulting impulse responses need not be similar. An IRF for a specific economic identification scheme will be a mixture of several IRFs from the reference rotation, where the linear combination is defined by the rotation matrix, $R$, that translates the baseline shocks into economically identified shocks. Even a small rotation can lead to different conclusions about the impact of structural shocks.

\subsection{Proximity to Perfect Correlation}

\begin{figure}
\centering{}\includegraphics[width=0.87\textwidth]{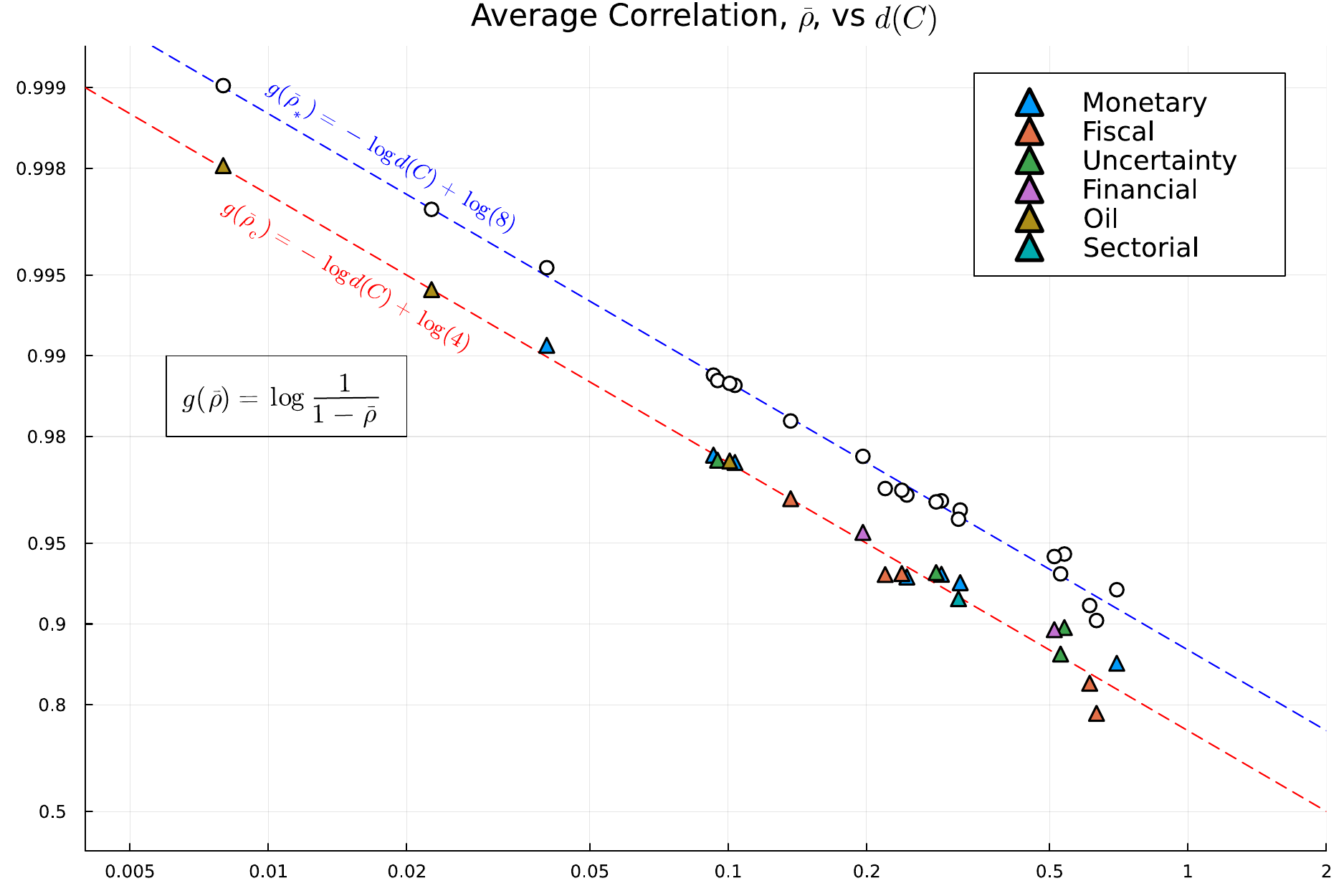}
\caption{{\small{}Scatterplot of $-\log(1-\bar\rho)$ against $\log d(C)$. The values shown along the axes are $\bar\rho$ and $d(C)$. OASIS points lie near the line $y=-x+\log 8$, whereas Cholesky points lie near $y=-x+\log 4$.\label{fig:CorrGap}}}
\end{figure}

The scatterplot in Figure \ref{fig:CorrGap} maps the proximity to perfect correlation, $-\log(1-\bar\rho)$, against the residual dependence in the reduced-form shocks, $\log d(C)$, for OASIS and Cholesky across all empirical applications. The dashed reference lines are the relationships predicted by (\ref{eq:ApproxOASIS}) and (\ref{eq:ApproxCholesky}) of Theorem~\ref{thm:Proximity}:
$
\bar\rho_{\ast}=1-\tfrac{1}{8}d(C)+O(\operatorname{tr}\{E^3\})$ and $
\bar\rho_{\mathrm{c}}=1-\tfrac{1}{4}d(C)+O(\operatorname{tr}\{E^3\})
$, respectively, with
$E=C-I_n$. Abstracting from the $O(\operatorname{tr}\{E^3\})$ terms, these become straight lines in a $(\log d(C),-\log(1-\bar\rho))$ plot with slope $-1$ and vertical intercepts $\log 8$ for OASIS and $\log 4$  for Cholesky. 

Empirically, the points fall tightly around these lines: OASIS observations align near
$$-\log(1-\bar\rho)\approx -\log d(C)+\log 8,$$
while Cholesky observations align near
$$-\log(1-\bar\rho)\approx -\log d(C)+\log 4.$$
This visualization makes the ``factor-of-two'' result immediate: the vertical separation $\log 2$ is equivalent to $(1-\bar\rho_{\mathrm{c}})\approx 2(1-\bar\rho_{\ast})$ in levels.

The interpretation is straightforward. The horizontal axis ($d(C)=\tfrac{1}{n}\|C-I\|_F^2$) measures how far the residuals are from being uncorrelated, such that a larger $d(C)$ means stronger contemporaneous comovement in reduced-form shocks. 
The vertical axis measures how close the structural shocks under each rotation are to their corresponding reduced-form innovations. The nearly linear log-log relationship, with common slope $-1$, shows that both schemes degrade at the same \emph{rate} as residual dependence rises, but OASIS is uniformly closer to perfect correlation as implied by theory. 

Deviations from the dashed lines are modest and attributable to the higher-order remainder terms $O(\operatorname{tr}\{E^3\})$ in the expansions.
Importantly, the choice of variable ordering influences this term, but the first-order expression $\bar\rho_{\mathrm{c}}=1-\tfrac{1}{4}d(C)+O(\operatorname{tr}\{E^3\})$ holds regardless, and any choice of variable ordering will land near the same location in this figure.  We observed that the values of $\bar\rho_\mathrm{c}$ do not systematically lie above or below the dashed line for Cholesky, so it does not appear that the orderings in the empirical literature were selected with this in mind.
By construction, OASIS is order-invariant, so each study contributes a single OASIS point, with no additional variation arising from the choice of variable ordering.

\subsection{Detailed Result for a Monetary Policy VAR}\label{sec:LSZ96detailed}

Next, we pursue a more detailed analysis of the SVAR used in \citet{Leeperetal1996} to study monetary shocks. We estimate a VAR(4) using quarterly U.S. data for the sample period  1959:Q1-2018:Q4. 
\citet{Leeperetal1996} used monthly data, which required them to interpolate quarterly GDP data to a monthly frequency.
The four variables are real GDP, the GDP deflator (DEF), the federal funds rate (FFR), and the money stock (M2). All variables are in log differences except the FFR, which is in levels.

The ordering of variables is (GDP, DEF, FFR, M2), which is the same as in \citet{Leeperetal1996}. 
We estimate the model with OASIS and the conventional lower-triangular Cholesky decomposition, as well as the upper-triangular Cholesky decomposition, which is identical to reversing the ordering of the variables.
Results for an alternative ordering and for a different price index are reported in the appendix.

The three approaches, the OASIS baseline and lower/upper Cholesky, provide three sets of IRFs. Here we focus on the IRFs for a monetary shock. 
\begin{table}[htbp]
\caption{OASIS and Cholesky applied to Leeper, Sims, and Zha (1996)}
\label{tab:LSZ1996}
{\setstretch{1.00} \footnotesize
\begin{tabularx}{\textwidth}{Xp{.2cm}
    >{\centering\arraybackslash}X
    >{\centering\arraybackslash}p{.2cm}
    >{\centering\arraybackslash}X
    >{\centering\arraybackslash}X
    >{\centering\arraybackslash}X
    >{\centering\arraybackslash}X
    >{\centering\arraybackslash}p{.2cm}
    >{\centering\arraybackslash}p{2.5cm}@{}}
\toprule

    && \multicolumn{1}{c}{ OASIS} &&  \multicolumn{4}{c}{ Cholesky} && \\
    &&       && Lower & Upper & $\min \bar\rho_\mathrm{c}$ & $\max \bar\rho_\mathrm{c}$ && \\[1mm]
\cmidrule{3-3}\cmidrule{5-8}\cmidrule{10-10}
    && $\operatorname{corr}(\varepsilon,u^\ast)$
        && \multicolumn{2}{c}{$\operatorname{corr}(\varepsilon,u^\mathrm{c})$}
        & \multicolumn{2}{c}{\footnotesize  Variable ordering}
        && $\operatorname{corr}(u^\ast,u^\mathrm{c})$    \\[1mm]
     GDP && 0.9995 && 1.0000 & 0.9974 & 3      & 1      && 0.9995 \\
     DEF && 0.9908 && 0.9999 & 0.9652 & 4      & 2      && 0.9908 \\
     FFR && 0.9751 && 0.9641 & 0.9356 & 2      & 4      && 0.9752 \\
     M2  && 0.9831 && 0.9343 & 1.0000 & 1      & 3      && 0.9836 \\[2mm] 
     \midrule
     \\[-4mm]
      \ \ $\bar\rho$  && 0.9871 && 0.9745 & 0.9745 & 0.9745 & 0.9750 && 0.9879 \\
    \bottomrule
\end{tabularx}
}
\end{table}

Table \ref{tab:LSZ1996} provides an example of how OASIS and Cholesky distribute the correlations across pairs of reduced-form shocks and structural shocks. The correlations are generally large. The individual correlations for OASIS are all above 97.5\% with an average of 98.71\%, whereas those for Cholesky range from 93.43\% to 100\% with an average value of 97.45\%. Table \ref{tab:LSZ1996} also reports the largest and smallest average correlation with Cholesky, and the corresponding variable orderings. There are twenty-four different ways to order the variables in this system. As predicted by Corollary \ref{cor:CholeskySimilar}, the average correlation is very similar across all variable orderings. In this application $\bar\rho_\mathrm{c}$ is within the narrow band between 97.45\% and 97.50\% for all Cholesky orderings. Note that the highest average Cholesky correlation is obtained by switching the order of the last two variables,  FFR and M2. Also observe that the correlations for the individual pairs $\operatorname{corr}(\varepsilon_j,u_j^\mathrm{c})$ vary substantially with the chosen ordering of the variables.

To see how structural shocks from a specific economic identification relate to those of the reference rotation, we can compute the rotation matrix $R$ defined in Proposition~\ref{prop:IRF}. The relationship between the Cholesky structural shocks, $u^c$, and the OASIS structural shocks, $u^\ast$, is given by $u^c = R^\prime u^\ast$ and $u^\ast = R u^c$, where $R =  A_\ast^{-1}A$, and in this empirical application we have
$$
R = \left[
\begin{array}{S[table-format=2.3] S[table-format=3.3] S[table-format=3.3] S[table-format=3.4]}
0.9995 & -0.0099 & -0.0227 & -0.0219 \\
0.0071 & 0.9908 & -0.1341 & 0.0160 \\
0.0273 & 0.1289 & 0.9752 & 0.1780 \\
0.0172 & -0.0397 & -0.1748 & 0.9836 \\
\end{array}
\right]
$$

The monetary shock is the third variable in $u^c$ and $u^\ast$, respectively, and the third column of the matrix $R$ tells us how the monetary shock baseline extracted by OASIS is ``relabeled'' as different structural shocks when identified by Cholesky.
For instance, the Cholesky structural shock to DEF is, in part, made up of the monetary shock baseline implied by OASIS, because $u_2^\mathrm{c}=0.1289 u_3^\ast +\cdots $. Similarly, the monetary shock identified by Cholesky, 
$$
u^c_3 = -0.0227 u^\ast_1 - 0.1341 u^\ast_2 + 0.9752 u^\ast_3 - 0.1748 u^\ast_4,
$$
has large components $-0.1341\,u_2^\ast$ and $-0.1748\,u_4^\ast$ that are interpreted as negative shocks to DEF and M2, respectively, in the OASIS baseline. 
This sheds light on discrepancies observed in the IRFs obtained with different identification restrictions. 
A structural shock, identified via economic restrictions, is a convolution of multiple benchmark shocks from the statistical reference rotation. The exact relations between the two types of shocks are given from the $R$-matrix. 

Cholesky's lower average correlation, $\bar\rho_{\mathrm{c}}<\bar\rho_{\ast}$, reflects a systematic down-scaling of the eigenvalue contributions. 
From the proof of Corollary~\ref{cor:OASIS}, the OASIS term is 
$\rho(A_\ast)=\sum_{i=1}^{n}\lambda_{i}^{1/2}$, 
whereas under Cholesky it becomes
$\rho(A_{\mathrm{c}})=\sum_{i=1}^{n} M_{ii}\,\lambda_{i}^{1/2}$,
with $\lambda_{1},\ldots,\lambda_{n}$ the eigenvalues of $C$ and 
$M=Q^\prime R^\prime Q$, where $C=Q\Lambda_\lambda Q^\prime$.
Since $M$ is orthonormal, $|M_{ii}|\le 1$ for all $i$, so the diagonal weights down-scale the eigenvalue square roots, yielding $\rho(A_{\mathrm{c}})\le \rho(A_\ast)$.
In this application we have
$$
M =Q^\prime R^\prime Q= \left[
\begin{array}{S[table-format=2.3] S[table-format=3.3] S[table-format=3.3] S[table-format=3.4]}
 0.9870 & -0.1360 & 0.0653 &  0.0555 \\
 0.1256 &  0.9804 & 0.0153 &  0.1510 \\
-0.0625 &  0.0013 & 0.9964 & -0.0574 \\
-0.0785 & -0.1425 & 0.0520 &  0.9853 \\
\end{array}
\right]\quad\text{and}\quad 
\left[ 
\begin{array}{c}
\sqrt{\lambda_1}\\
\sqrt{\lambda_2}\\
\sqrt{\lambda_3}\\
\sqrt{\lambda_4}\\
\end{array}
\right]
=
\left[ 
\begin{array}{S[table-format=2.3] }
0.774 \\
0.947 \\
1.006 \\
1.222 \\
\end{array}
\right].
$$
The diagonal elements of $M$ show how much the square roots of the eigenvalues of $C$ get scaled down by Cholesky, which is the reason Cholesky has a smaller average correlation.

\subsection{Empirical VARs with more reduced-form correlation}

All the studies in Table~\ref{tab:shock_studies} have reduced-form innovations with relatively low cross-correlations, as indicated by the small $\|C\|_1^{\star}$ and $d(C)=\tfrac{1}{n}\|C-I\|_{F}^{2}$ reported in Table~\ref{tab:EmpiricalResults}. When $d(C)$ is small, OASIS and Cholesky both deliver structural shocks that align closely with their labeled reduced-form innovations.

To explore cases where the methods diverge more, we included three VARs with higher residual correlation: two drawn from existing work, and a third (a term-structure VAR) we constructed.
\begin{table}[htbp!]
\caption{Empirical VARs with More Correlations in Residuals}
\label{tab:shock_studies2}
{\setstretch{1.00}\footnotesize
    \begin{tabularx}{\textwidth}{@{}p{1.45cm}X @{}}
        \toprule
        \multicolumn{2}{@{}l}{\textbf{Topic/[Abbreviation]/Reference/Notes}} \\[1mm]
        \midrule
        \multicolumn{2}{c}{\textbf{VARs with Correlated Residuals}}\\[2mm]
        {[EI95]} &  \citet{EngleIssler:1995} VECM testing long-run relationship of U.S. sectoral output data  \\
        {[DP05]} &  \citet{DaiPhilippon:2005} VAR identifying fiscal shocks via structural budget components; focuses on term structure. \\
        {[FHT25]} &  (This paper) A term structure VAR, see description in Section \ref{TSVAR} for details. \\
        \bottomrule
    \end{tabularx}
    \vspace{0.3mm}\\   
    \noindent\textit{Notes:} See notes for Table \ref{tab:shock_studies} \\
}
\end{table}

The first of these studies is \cite{EngleIssler:1995}. They did not use Cholesky, but did analyze productivity shocks, and their VAR has correlated residuals that, in part, are driven by input-output linkages across sectors. This VAR is therefore well suited for a comparison of OASIS and Cholesky in a setting with high residual correlations.  \cite{EngleIssler:1995} build on the Real Business Cycle model of \cite{longplosser1983} and show that cointegration among sectoral outputs implies cointegration among underlying productivity shocks, linking long-run comovement to technological fundamentals. Using a VECM and U.S. sectoral per-capita output data (1947-1989), they identify two cointegrating (long-run) vectors and six cofeature (shorter-run) vectors.

The second study is \cite{DaiPhilippon:2005} (an unpublished NBER working paper) that estimates a VAR-based affine term structure model using quarterly data from 1970:Q1 to 2003:Q3. The model incorporates eight Treasury bond yields within a no-arbitrage affine term structure framework to study the effects of fiscal policy shocks on interest rates. The authors use a VAR for macroeconomic variables and latent yield factors, while the eight bond yields enter the model through the measurement equation. Structural fiscal shocks are identified using a recursive identification scheme similar to \cite{blanchard2002}. The authors then decompose the response of long-term yields into changes in expected future short rates and term premia. They find that sustained increases in fiscal deficits raise long-term interest rates over time.

\label{TSVAR}
The third study is a term structure VAR, with seven yields spanning a wide range of maturities, from the overnight federal funds rate to the 10-year Treasury Bill (T-Bill) yield, as well as intermediate T-Bill yield at the 3-month, 6-month, 1-year, 3-year, and 5-year horizons. We estimate a VAR(12) using monthly data from September 1981 to January 2025.

\begin{table}[htbp!]
\caption{OASIS and Cholesky in Highly Correlated VARs}
\label{tab:EmpiricalResults2}
{\setstretch{1.00}\footnotesize
\begin{tabularx}{\textwidth}{@{}>{\hsize=1.5\hsize}X
>{\hsize=0.25\hsize}X 
>{\hsize=1.75\hsize\centering\arraybackslash}X
>{\hsize=0.75\hsize\centering\arraybackslash}X
>{\hsize=0.75\hsize\centering\arraybackslash}X
>{\hsize=0.75\hsize\centering\arraybackslash}X
>{\hsize=1.75\hsize\centering\arraybackslash}X
>{\hsize=0.75\hsize\centering\arraybackslash}X
>{\hsize=1.00\hsize\centering\arraybackslash}X
>{\hsize=0.75\hsize\centering\arraybackslash}X@{}}
\toprule
 &  & \multicolumn{1}{c}{OASIS} & \multicolumn{3}{c}{Cholesky} &  &  & &  \\
Study & 
$n$ & \multicolumn{1}{c}{$\bar{\rho}_\ast$} & \multicolumn{1}{c}{$\bar{\rho}_\mathrm{c}$} & min & max & $\bar{\rho}_{\ast,\mathrm{c}}$  & $\|C\|_1^{\star}$  & $\tfrac{1-\bar{\rho}_\mathrm{c}}{1-\bar{\rho}_\ast}$ & $d(C)$\\[3mm]
\midrule
    \\[-2mm]
     {[EI95]} & 8 & 0.821 & 0.704 & 0.695 & 0.718   & 0.867 & 0.47 & 1.66& 2.00  \\[1mm]
     {[DP05]} & 8 & 0.621 & 0.345 & 0.301 & 0.407   & 0.477 & 0.64 & 1.73& 3.29  \\[1mm]
     {[FHT25]} & 7 & 0.684 & 0.461 & 0.456 & 0.516  & 0.635 & 0.69 & 1.71& 3.12  \\[2mm]
\bottomrule
\end{tabularx}
\vspace{0.3em}\\
\noindent\textit{Notes:} See notes for Table \ref{tab:EmpiricalResults}.
}
\end{table}

The three VARs estimated in Table \ref{tab:EmpiricalResults2} have residuals with substantially more correlations than those in  Table \ref{tab:EmpiricalResults}. This can be seen from the higher values of $\|C\|_1^{\star}$  and $d(C)$. Specifically, the highest values in Table \ref{tab:EmpiricalResults} are $\|C\|_1^{\star} = 0.26$ and $d(C) = 0.70$, whereas the lowest values in Table \ref{tab:EmpiricalResults2} are $\|C\|_1^{\star} = 0.47$ and $d(C) = 2.00$. Consequently, the average correlations, $\bar{\rho}_\ast$ and $\bar{\rho}_\mathrm{c}$, are smaller and the difference between the two is larger,  as predicted by Theorem \ref{thm:Proximity}. The ratios, $\tfrac{1-\bar{\rho}_\mathrm{c}}{1-\bar{\rho}_\ast}$, deviate further from 2 because the third order term, $O(\operatorname{tr}\{(C-I)^{3}\})$, is larger for these three VARs.

\section{Revisiting Proxy-VAR and SVAR-IV Studies}\label{sec:ProxyApplications}

In this section, we revisit two seminal applications to demonstrate how OASIS can be applied to the Proxy-VAR framework, and how it can be used to estimate leakage and measure instrument strength. 

We first revisit \citet{MertensRavn:2013}, which utilizes a single proxy variable per structural shock.
This falls within the structure analyzed in Section 3.1.
We then revisit \citet{StockWatson:2012}, which utilizes multiple proxy variables for each structural shock. This
falls within the identification framework analyzed in Section 3.2 and we can employ the test for overidentification we developed in Section 3.3.

These exercises demonstrate the empirical necessity of accommodating structural leakage. In the exactly identified setting of \citet{MertensRavn:2013}, we show that replacing their asymmetric recursive (Cholesky) restriction with our symmetric proportional leakage framework reveals severe cross-shock contamination, which meaningfully alters the estimated macroeconomic multipliers. In the overidentified setting of \citet{StockWatson:2012}, we are able to formally test and strongly reject the classical assumption of strict proxy orthogonality. Together, these applications highlight the empirical prevalence of proportional leakage and the practical value of the OASIS rotation.

\subsection{Revisiting Narrative Tax Shocks (Mertens and Ravn, 2013)}

We revisit the highly influential Proxy-SVAR analysis of US tax policy by \citet{MertensRavn:2013}. The authors utilize narrative records of legislative tax changes to construct two distinct proxy variables: one targeting Personal Income (PI) tax shocks, and another targeting Corporate Income (CI) tax shocks. 

Because these proxies are empirically correlated, \citet{MertensRavn:2013} explicitly acknowledge that they suffer from cross-contamination. To achieve exact identification within the standard Proxy-SVAR framework, they impose an asymmetric recursive ordering (a Cholesky decomposition) between the instruments. Mathematically, this amounts to a strict zero-restriction: it assumes one proxy is perfectly ``clean'' of the alternative structural shock, forcing the second proxy to absorb all the residual correlation. 

This application can be embedded in the two-shock proxy setting developed in Section 3; under symmetric proportional leakage, the observable $H$ matrix admits closed-form recovery of the contamination parameter.

First, we partition the $9 \times 9$ sample covariance matrix provided by the dataset, standardize the narrative proxies to unit variance, and compute the $2 \times 2$ observable signal matrix $H \equiv \Sigma_{z\varepsilon}\Sigma_{\varepsilon\varepsilon}^{-1}\Sigma_{\varepsilon z}$. This yields:
$$
H = \begin{bmatrix} 
0.0635 & 0.0191 \\ 
0.0191 & 0.0294 
\end{bmatrix}
$$
Under the assumption of symmetric proportional leakage, the estimated cross-loading is $\hat{s}_{1,2}=0.2336$, indicating economically meaningful contamination of each tax proxy by the other tax shock. The recursive Proxy-SVAR framework accommodates this dependence asymmetrically by imposing a triangular structure, effectively assigning the residual cross-correlation to one proxy rather than the other. By contrast, the OASIS rotation treats this contamination symmetrically and yields implied own-shock signal strengths of  $\hat{\alpha}_{\text{PI}} = 0.2453$ and $\hat{\alpha}_{\text{CI}} = 0.1669$.

The asymmetric Cholesky restriction can materially shape the estimated macroeconomic responses. When one proxy is treated as clean and the remaining cross-correlation is absorbed by the other, the resulting impulse responses need not isolate the two tax shocks cleanly, and may instead reflect a mixture of their effects.

As established in Section 3, the squared values of these purified signal strengths ($\hat{\alpha}_{\text{PI}}^2 \approx 0.060$ and $\hat{\alpha}_{\text{CI}}^2 \approx 0.028$) can be interpreted as measures of structural first-stage relevance. These values indicate that the instrument strength of the CI proxy is weaker than the PI proxy.

Table \ref{tab:MertensRavn_Multipliers} reports the contemporaneous impact of a 1 percentage point cut in either PI or CI taxes on Real GDP and Federal Debt. We compare the classical Proxy-SVAR estimates against the robust OASIS estimates, which optimally symmetrize the signal matrix by setting $\Lambda_w = \Lambda_\alpha^{-1}$ to partial out the $\hat{s} = 0.2336$ leakage.

\begin{table}[htbp]
    \caption{Contemporaneous Macroeconomic Impact of a 1 pp Tax Cut}
    \label{tab:MertensRavn_Multipliers}
    {\setstretch{1.00}\footnotesize
    \begin{tabularx}{\textwidth}{X >{\centering\arraybackslash}X >{\centering\arraybackslash}X c >{\centering\arraybackslash}X >{\centering\arraybackslash}X}
        \toprule
        & \multicolumn{2}{c}{\textbf{Real GDP}} & & \multicolumn{2}{c}{\textbf{Federal Debt}} \\
        \cmidrule(lr){2-3} \cmidrule(lr){5-6}
        \textbf{Shock} & Recursive Proxy-VAR & OASIS & & Recursive Proxy-VAR & OASIS \\
        \midrule
        PI Tax Cut & 0.827 & 0.582 & & 0.531 & 0.510 \\
        CI Tax Cut & 0.951 & 0.638 & & 0.212 & 0.094 \\
        \bottomrule
    \end{tabularx}
     \vspace{0.3mm}\\   
     \noindent\textit{Notes:} {We report the estimated contemporaneous response of Real GDP and Federal Debt to a 1 percentage point cut in Personal Income (PI) and Corporate Income (CI) taxes. The Recursive Proxy-VAR estimates isolate the shocks by imposing a recursive (Cholesky) ordering between the tax instruments. The OASIS estimates instead explicitly account for the estimated $\hat{s} = 0.2336$ cross-shock contamination symmetrically, without requiring an arbitrary ordering.} 
    }
\end{table}

Accounting for proxy leakage symmetrically leads to materially different estimates of the contemporaneous macroeconomic responses. Relative to the recursive Proxy-SVAR specification, the OASIS rotation yields smaller estimated effects of both PI and CI tax cuts. For PI taxes, the contemporaneous response of Real GDP declines from $0.827$ to $0.582$, a reduction of roughly 30 percent. For CI taxes, the corresponding Real GDP response declines from $0.951$ to $0.638$, while the estimated response of Federal Debt falls from $0.212$ to $0.094$. These differences suggest that resolving proxy cross-correlation asymmetrically can affect the estimated responses in economically meaningful ways, whereas the OASIS framework provides a symmetric alternative that does not rely on a chosen recursive ordering.

\subsection{Channels of the Great Recession (Stock and Watson, 2012)}

The standard SVAR-IV framework employed by \citet{StockWatson:2012} relies on the assumption of strict instrument exogeneity, which dictates that the narrative proxies for a given shock are completely uncorrelated with all other structural shocks in the system. In the context of our proportional leakage framework (Assumption 1$^{\prime\prime}$), this strict assignment condition requires that $s_{ij} = 0$ for all $i \neq j$, such that the structural leakage matrix $S$ is perfectly diagonal. 

We can directly test this implicit assumption by applying the OASIS framework to their dataset. Applying the identification framework of Theorem 6 yields the following estimate of the structural leakage matrix $S$.

{\begin{small}
\setlength{\abovedisplayskip}{0pt}
$$
\hat{S} = \left[\begin{array}{cccccc}
1 & \phantom{-}0.116^{\ast\ast\ast} & -0.324^{\ast\ast\ast} & -0.457^{\ast\ast\ast} & -0.219^{\ast\ast\ast} & \phantom{-}0.104^{\ast\ast\ast} \\
\phantom{-}0.116^{\ast\ast\ast} & 1 & \phantom{-}0.218^{\ast\ast\ast} & \phantom{-}0.023\phantom{^{\ast\ast\ast}} & -0.175^{\ast\ast\ast} & \phantom{-}0.413^{\ast\ast\ast} \\
-0.324^{\ast\ast\ast} & \phantom{-}0.218^{\ast\ast\ast} & 1 & -0.007\phantom{^{\ast\ast\ast}} & \phantom{-}0.043^{\ast\ast\ast} & \phantom{-}0.201^{\ast\ast\ast} \\
-0.457^{\ast\ast\ast} & \phantom{-}0.023\phantom{^{\ast\ast\ast}} & -0.007\phantom{^{\ast\ast\ast}} & 1 & \phantom{-}0.661^{\ast\ast\ast} & -0.071^{\ast\ast\ast} \\
-0.219^{\ast\ast\ast} & -0.175^{\ast\ast\ast} & \phantom{-}0.043^{\ast\ast\ast} & \phantom{-}0.661^{\ast\ast\ast} & 1 & \phantom{-}0.087^{\ast\ast\ast} \\
\phantom{-}0.104^{\ast\ast\ast} & \phantom{-}0.413^{\ast\ast\ast} & \phantom{-}0.201^{\ast\ast\ast} & -0.071^{\ast\ast\ast} & \phantom{-}0.087^{\ast\ast\ast} & 1
\end{array}\right]
\quad
\left[\begin{array}{c}
\text{Oil}\\
\text{Monetary}\\
\text{Productivity}\\
\text{Uncertainty}\\
\text{Liq./Finan.}\\
\text{Fiscal}
\end{array}\right]
$$
\end{small}
}

As shown below, the robust rank test broadly supports the proportional-leakage structure, with the Uncertainty block providing the only rejection.

The estimated structural leakage matrix, $\hat{S}$, reveals pervasive cross-shock dependencies among the proxy variables. Using wild bootstrap standard errors, we find that nearly all off-diagonal elements are statistically significant at the 1\% level. This overwhelmingly rejects the classical exclusion restriction underlying the baseline case.

The magnitudes and signs of these leakages align closely with economic intuition. For example, measures of uncertainty are naturally responsive to both financial risk and energy-market disruptions, which is reflected in the large cross-loadings between the Uncertainty composite proxy and the Liquidity/Financial structural shock ($\hat{s}_{45}=0.661$), and between the Uncertainty composite proxy and the Oil structural shock ($\hat{s}_{14}=-0.457$). The negative sign indicates that, under our normalization, the uncertainty proxy loads negatively on the oil shock, consistent with episodes in which adverse uncertainty events are associated with falling oil prices. In fact, the only statistically insignificant cross-loadings in the entire system occur between Uncertainty and the Monetary and Productivity shocks. Similarly, the narrative Monetary and Fiscal Policy proxies exhibit a substantial spillover ($\hat{s}_{26}=0.413$), reflecting the fact that these proxy blocks respond to closely related macroeconomic disturbances.

Because these proxies are heavily cross-contaminated, assuming a diagonal structure for the composite signal matrix $\bar{\Phi}$ is empirically invalid. Attempting to force these off-diagonal correlations to exactly zero strongly conflicts with the data and would severely distort the identified structural rotation. By contrast, the OASIS framework accommodates this empirical reality. By optimally symmetrizing the cross-correlation matrix, OASIS explicitly accounts for severe leakages, such as the $0.661$ spillover between Uncertainty and Liquidity, allowing for robust point identification without requiring theoretically unjustified zero-restrictions.

\begin{table}[htbp]
    \caption{Signal Strength, Leakage, and Test for Overidentification}
    \label{tab:rank_tests}
    {\setstretch{1.00}\footnotesize
    \begin{tabularx}{\textwidth}{
        X 
        >{\hsize=0.8\hsize\centering\arraybackslash}X 
        >{\hsize=0.8\hsize\centering\arraybackslash}X 
        >{\hsize=1.1\hsize\centering\arraybackslash}X 
        >{\hsize=1.3\hsize\centering\arraybackslash}X
    }
        \toprule
        & & & \multicolumn{2}{c}{Rank tests (Statistics)} \\
        \cmidrule(lr){4-5}
        Shock Type & $\alpha_i^2$ & $s_{i\bullet}^2$ & Anderson & Kleibergen-Paap \\
        \midrule
        Oil                 &  0.198    & 1.386 & \textcolor{darkgray}{10.63\phantom{$^{***}$}}     &  \phantom{0}7.49\phantom{$^{***}$} \\
        Monetary            &  0.445    & 1.262 & \textcolor{darkgray}{15.04\phantom{$^{***}$}}     & 41.14$^{***}$ \\
        Productivity        &  0.237    & 1.195 & \textcolor{darkgray}{12.45$^{**}$\phantom{$^*$}}  &  \phantom{0}5.57\phantom{$^{***}$} \\
        Uncertainty         &  0.364    & 1.651 & \textcolor{darkgray}{34.92$^{***}$}               & 23.80$^{***}$ \\
        Liquidity/Financial &  0.320    & 1.524 & \textcolor{darkgray}{13.52\phantom{$^{***}$}}     &  \phantom{0}2.75\phantom{$^{***}$} \\
        Fiscal              &  0.041    & 1.234 & \textcolor{darkgray}{ \phantom{0}3.02\phantom{$^{***}$}}     &  \phantom{0}1.86\phantom{$^{***}$} \\
        \bottomrule
    \end{tabularx}
    \vspace{0.3mm}\\   
    \noindent\textit{Notes:} {This table reports the signal strength ($\alpha_i^2$), the leakage multiplier ($s_{i\bullet}^2$), and the test statistics for the null hypothesis of proportional leakage (rank=1). The Anderson trace statistic assumes homoskedasticity, while the Kleibergen-Paap $J_\mathrm{KP}$-statistic is robust to heteroskedasticity and autocorrelation. Significance levels: $^{*} p < 0.10$, $^{**} p < 0.05$, $^{***} p < 0.01$.}
    }
\end{table}

Table \ref{tab:rank_tests} reports the estimated signal strength ($\alpha_i^2$) and the leakage multiplier ($s_{i\bullet}^2$) for each structural shock, alongside the formal tests for overidentification. The signal strengths reveal considerable heterogeneity across the instruments, with the Monetary policy proxies exhibiting the strongest structural relevance ($\alpha_i^2 = 0.445$) and the Fiscal proxies remaining notably weak ($\alpha_i^2 = 0.041$). The leakage multipliers, which capture the variance inflation caused by cross-shock contamination ($s_{i\bullet}^2 \geq 1$), further quantify the proxy spillovers discussed above. The Uncertainty block suffers the most severe inflation ($s_{i\bullet}^2 = 1.651$), meaning that the variance of the proxy block is inflated by 65 percent due to spillovers from other macroeconomic shocks rather than pure uncertainty innovations.

Next, we apply the test for overidentification. The right panel of Table \ref{tab:rank_tests} presents the test statistics for the null hypothesis of proportional leakage, which restricts the relevant cross-covariance matrix to a rank of exactly one. The classical Anderson trace statistic assumes conditional homoskedasticity, whereas the robust Kleibergen-Paap $J_\mathrm{KP}$-statistic accounts for both heteroskedasticity and serial correlation. As the table illustrates, explicitly accounting for these features can meaningfully alter the empirical conclusions.

The robust Kleibergen-Paap test fails to reject the null hypothesis for four of the six shock blocks. This provides partial support for the proportional leakage assumption (Assumption 1$^{\prime\prime}$) in these cases. It confirms that the multiple proxy variables within the Oil, Productivity, Liquidity/Financial, and Fiscal blocks share collinear structural leakage profiles. Therefore, they can be validly collapsed into their respective composite instruments without distorting the underlying structural identification. 

The two exceptions are the Uncertainty and Monetary blocks, which both firmly reject the proportional leakage restriction ($p < 0.01$). For the Uncertainty proxies, this rejection indicates that the underlying measures, such as the VIX and the Economic Policy Uncertainty index, capture fundamentally different structural dynamics rather than parallel measurements of a single underlying shock. Consequently, collapsing these specific proxies into a single scalar instrument via $\beta_i^\ast$ implies a loss of distinct structural information. For the Monetary block, the strong rejection by the robust test should be interpreted with some caution. The available data for these specific joint instruments yields a highly restricted effective sample size of just $T=28$, largely because the \citet{GurkaynakSackSwanson:2005} series begins in 1990 while the \citet{RomerRomer:2004} series ends in 1996. Because HAC estimators can be unreliable in such small samples, the rejection may reflect finite-sample distortion rather than a genuine violation of proportional leakage.

\section{Summary and Concluding Remarks}\label{sec:Summary}

In this paper, we have proposed a different perspective on identification in dynamic macroeconomic models. Instead of relying solely on hard restrictions, the mapping between structural shocks and designated target variables can be characterized as the solution to an explicit optimization problem. The maximum-correlation criterion selects the orthogonal rotation that maximizes the average correlation between structural shocks and their targets, leading to the Order- and Scale-Invariant Scheme (OASIS).

By reinterpreting the zero restrictions of a Cholesky decomposition as the first-order conditions of a sequential optimization problem, we show that recursive identification implicitly solves a constrained version of the maximum-correlation objective. Their shared objective helps explain why various Cholesky schemes tend to produce broadly similar impulse responses. Furthermore, because the aggregate correlation is nearly identical across all orderings, the choice of causal ordering effectively reduces to a choice of how to distribute that correlation across shocks. By explicitly maximizing this correlation without recursive constraints, OASIS provides a unique, order- and scale-invariant orthogonal rotation. This makes it an ideal diagnostic baseline when the causal ordering is ambiguous. Our theoretical results are strongly supported by the twenty-two empirical SVAR studies we revisited.

Importantly, beyond its role as a diagnostic baseline for internal variables, the OASIS framework provides a coherent estimation strategy for Proxy VARs (IV-SVARs). As established in Section \ref{sec:ProxyVAR}, by maximizing the aligned correlations between structural shocks and external instruments, OASIS uses instrument relevance as the core estimation criterion. It provides a principled route to point identification that resolves overdetermination in multi-shock settings, bypassing zero-covariance restrictions that may be difficult to justify or inconsistent with the data. By symmetrically accommodating proxy-shock correlations rather than arbitrarily restricting them to zero, OASIS preserves the orthonormal structure of the shocks while naturally handling proxy leakage. As demonstrated in Section \ref{sec:ProxyApplications}, applying this framework to seminal empirical models reveals that formally accounting for such leakage can materially affect empirical conclusions in important applications.

OASIS is also particularly complementary to sign-restricted identification. Standard sign restrictions yield a set of admissible models rather than a point estimate, leading to well-known difficulties in interpretation and inference \citep{FryPagan:2011}. By maximizing the OASIS objective subject to sign constraints, researchers can select a single, economically plausible benchmark model from the admissible set using a transparent statistical criterion. This provides a computationally straightforward approach that helps mitigate the unintended prior distortions often introduced by standard Bayesian sampling algorithms \citep{AriasRubioRamirezWaggoner:2018}.

By shifting the evaluation of identification from hard restrictions to an objective function, we shift the focus of the discussion: the question is no longer just whether the zero restrictions are reasonable, but how the resulting rotation compares to a statistical baseline. While definitive structural identification ultimately requires exogenous economic insight, the maximum-correlation criterion provides a transparent benchmark for assessing empirically dubious identification schemes. Although many observationally equivalent rotations are consistent with the same reduced-form covariance matrix, many such matrices are difficult to defend because they imply structural shocks that are essentially unrelated to the key variables with which they should be aligned. For example, labeling a structural shock as ``monetary'' when it is uncorrelated with standard monetary variables, or deploying a narrative tax instrument that exhibits low relevance for its targeted shock, is economically counterintuitive. The OASIS benchmark naturally avoids these pathological outcomes by explicitly anchoring structural shocks to their designated targets. 

Future research could explore extensions of OASIS to settings where an order- and scale-invariant reference rotation is particularly beneficial. One natural extension is to models with time-varying volatility or heteroskedasticity \citep{ChanKoopYu:2024-JBES, GoncalvesHerreraKilianPesavento:2024}, where OASIS can be applied sequentially to retain order-neutral likelihoods while enforcing a stable structural representation. Another promising application is impulse response function (IRF) matching \citep[e.g.,][]{basu2017}, where theoretically implied structural shocks from a DSGE model can serve as target directions for extracting the corresponding empirical shock combinations from VAR innovations.

\pagebreak

\section*{Data and Code Availability Statement}

The replication package for this paper, including all data, code, and intermediate outputs necessary to replicate the tables, figures, and empirical findings, is publicly available on GitHub at \url{https://github.com/reinhardhansen/OASIS-Replication}. 

\paragraph{Software Environment}
The primary empirical analyses, simulations, and optimizations were conducted using Julia (v1.12+). The core matrix operations and optimizations rely heavily on the standard \texttt{LinearAlgebra} library. To ensure the robustness of the empirical covariance matrices, the initial data processing and covariance estimations were performed in \textbf{Matlab} (\texttt{CovmatrixMatlab.m}) and cross-verified using \textbf{R} (\texttt{CovmatrixRcode.R}). Final computations are executed via a provided Jupyter Notebook (\texttt{SVARidentification.ipynb}). 

\paragraph{Data Provenance}
The empirical analysis relies on three distinct sets of data:
\begin{enumerate}
    \item \textbf{The 22 SVARs Revisit:} The reduced-form covariance matrices and macroeconomic time series for the studies listed in Table \ref{tab:shock_studies} and Table \ref{tab:shock_studies2} were obtained directly from the original authors' published replication packages. 
    \item \textbf{Monetary Policy VAR (Section \ref{sec:LSZ96detailed}):} The quarterly U.S. data (1959:Q1--2018:Q4) for real GDP, the GDP deflator, the federal funds rate, and M2 were sourced from the Federal Reserve Economic Data (FRED) database. 
    \item \textbf{Term Structure VAR (Section \ref{TSVAR}):} The monthly yield curve data spanning September 1981 to January 2025 were sourced from the Federal Reserve Board's nominal yield curve estimates. 
\end{enumerate}
Complete details regarding variable transformations, exact sample periods, and FRED series identifiers (e.g., \texttt{FEDFUNDS.csv}, \texttt{GDPC1.csv}) are documented in Appendix B and the \texttt{README.md} file included in the GitHub repository.

\bibliographystyle{apalike}
\bibliography{oasis,prh}

\newpage
\appendix
\setcounter{equation}{0}
\global\long\def\theequation{A.\arabic{equation}}%
\setcounter{lemma}{0}
\global\long\def\thelemma{A.\arabic{lemma}}%
\section{Appendix of Proofs}

\noindent {\bf Proof of Proposition \ref{prop:Chol}.}
Let $\Sigma_{\varepsilon\varepsilon} = LL^\prime$ be the Cholesky decomposition of the reduced-form covariance matrix, where $L$ is lower triangular with strictly positive diagonal entries. Under recursive identification, the structural matrix is $A_c^\prime = L^{-1}$.

At step $j$, the problem is to choose $a_j$ to maximize
$$
\operatorname{corr}(a^\prime \varepsilon,\varepsilon_j)
\qquad\text{s.t.}\quad
a^\prime \Sigma_{\varepsilon\varepsilon} a = 1
\text{ and }
a^\prime \Sigma_{\varepsilon\varepsilon} a_i = 0
\quad\text{for all } i<j.
$$

Because $\operatorname{var}(a^\prime\varepsilon)=1$, we have
$$
\operatorname{corr}(a^\prime \varepsilon,\varepsilon_j)
=
\frac{\operatorname{cov}(a^\prime \varepsilon,\varepsilon_j)}
{\sigma_{\varepsilon_j}}
=
\frac{a^\prime \Sigma_{\varepsilon\varepsilon} e_j}{\sigma_{\varepsilon_j}},
$$
so maximizing the correlation is equivalent to maximizing
$a^\prime \Sigma_{\varepsilon\varepsilon} e_j
=a^\prime LL^\prime e_j
$.
If we define
$x\equiv L^\prime a$.
Then
$a^\prime \Sigma_{\varepsilon\varepsilon} a
=a^\prime LL^\prime a=x^\prime x=1$,
expresses the unit-variance constraint, and the objective becomes
$a^\prime LL^\prime e_j = x^\prime L^\prime e_j$.

We now prove by induction that the maximizer at step $j$ is
$a_j = L^{-1\prime} e_j$.

For $j=1$, there are no orthogonality constraints. Since $L^\prime e_1$ has only its first component nonzero, the objective is proportional to $x_1$, and under $x^\prime x=1$ it is uniquely maximized by $x=e_1$. Hence
$\hat{a}_1=(L^\prime)^{-1}e_1$. 
Next, consider the case for some $j\ge 2$, where the orthogonality constraints become
$$
0=a^\prime \Sigma_{\varepsilon\varepsilon} a_i
=a^\prime LL^\prime (L^\prime)^{-1}e_i
=a^\prime Le_i
=x^\prime e_i
=x_i,
\qquad i<j,
$$
which shows that any feasible $x$ satisfies,
$x_1=\cdots=x_{j-1}=0$.

Next, note that $L^\prime e_j$ is the $j$th column of $L^\prime$. Since $L$ is lower triangular, this vector has zero entries below position $j$. Therefore
$x^\prime L^\prime e_j = x_j L_{jj}$,
because the first $j-1$ components of $x$ are zero and the components of $L^\prime e_j$ after position $j$ are also zero.

Since $L_{jj}>0$, maximizing the objective is equivalent to maximizing $x_j$ subject to $x^\prime x=1$. The unique maximizer is
$x=e_j$ and therefore
$$
\hat{a}_j=(L^\prime)^{-1}e_j,\qquad \text{for}\quad j=1,\ldots,n.
$$
This proves that at every step $j$, the solution is the $j$-th column of $A_c=(L^{\prime})^{-1}$. Hence the sequential maximum-correlation problem reproduces the Cholesky identification.
\hfill$\square$

\noindent {\bf Proof of Theorem \ref{thm:WeightedOASIS}.} 
From $\operatorname{cov}(u,\varepsilon)=A^{\prime}\Sigma_{\varepsilon\varepsilon}=A^{\prime}\Lambda_{\sigma_\varepsilon} C_{\varepsilon\varepsilon} \Lambda_{\sigma_\varepsilon}$ we have $\operatorname{corr}(u,\varepsilon)=A^{\prime}\Lambda_{\sigma_\varepsilon} C_{\varepsilon\varepsilon}$, such that 
$$\rho_w(A)=\operatorname{tr}\{A^{\prime}\Lambda_{\sigma_\varepsilon} C_{\varepsilon\varepsilon}\Lambda_w\}
=\operatorname{tr}\{\check{A}^{\prime}\Lambda_wC_{\varepsilon\varepsilon}\Lambda_w\},\qquad \check{A}^{\prime}\equiv A^{\prime}\Lambda_{\sigma_\varepsilon} \Lambda_w^{-1}.$$ 
From the eigendecomposition, 
$\Lambda_w C_{\varepsilon\varepsilon}\Lambda_w=Q\Lambda_{\lambda(w)}Q^{\prime}$, we define $\check{A}_{\ast}^{\prime}=Q\Lambda_{\lambda(w)}^{-1/2}Q^{\prime}$ for which we have $\rho_w(A_\ast)=\operatorname{tr}(\Lambda_{\lambda(w)}^{1/2})=\sum_{i=1}^n \lambda_i^{1/2}(w)$ for $A_{\ast}^\prime=Q\Lambda_{\lambda(w)}^{-1/2}Q^{\prime}\Lambda_w\Lambda_{\sigma_\varepsilon}^{-1}$. It is straightforward to verify that $A_{\ast}\in\mathcal{A}$, and for any $A\in\mathcal{A}$, we have $A=A_\ast R$ for some orthonormal $R\in\mathbb{R}^{n\times n}$ (i.e., $R^\prime R=I_n$), so that
$$\rho_w(A)=\operatorname{tr}\{R^{\prime}A_\ast^{\prime}\Lambda_{\sigma_\varepsilon} C_{\varepsilon\varepsilon}\Lambda_w\}
=\operatorname{tr}\{R^{\prime} Q\Lambda_{\lambda(w)}^{1/2}Q^\prime\}
=\operatorname{tr}\{M\Lambda_{\lambda(w)}^{1/2}\}=\sum_{i=1}^n M_{ii}\lambda_i^{1/2}(w)\leq \sum_{i=1}^n \lambda_i^{1/2}(w),$$
where $M=Q^\prime R^{\prime} Q$ is an orthonormal matrix, with $M_{ii}\leq 1$ for all $i=1,\ldots,n$.

If $\rho_w(A)=\rho_w(A_\ast)$, then $M_{ii}=1$ for all $i$ and because $M$ is orthonormal, this implies $M=I_n$, hence it must be that $A=A_\ast$.
Finally, this identification scheme is order- and scale-invariant because the eigenvalues of $\Lambda_w C_{\varepsilon\varepsilon} \Lambda_w$ do not depend on the ordering of variables nor their scale. 
\hfill$\square$\medskip{}

Although Corollary \ref{cor:OASIS} is an immediate consequence of Theorem \ref{thm:WeightedOASIS}, we include a direct proof because it provides useful intuition for the equal-weights case and develops ideas that are referenced elsewhere in the paper.

\noindent {\bf Proof of Corollary \ref{cor:OASIS}.} 
We begin by solving a simpler auxiliary problem, where the
reduced-form shocks, denoted $\eta_{i}$, $i=1,\ldots,n$, have unit variances such that $\operatorname{var}(\eta)=C_{\eta\eta}$ is
a correlation matrix. Here, a vector of structural shocks is given
by $u=\tilde{A}^{\prime}\eta$, where $\tilde{A}$ satisfies $\tilde{A}^{\prime}C_{\eta\eta}\tilde{A}=I_n$. The aggregate
correlation between $u_{i}$ and $\eta_{i}$ is given by
\[
\varrho(\tilde{A})\equiv\sum_{i=1}^{n}\operatorname{corr}(u_{i},\eta_{i})=\operatorname{tr}\{\operatorname{cov}(\tilde{A}^{\prime}\eta,\eta)\}=\operatorname{tr}\{\tilde{A}^{\prime}C_{\eta\eta}\}.
\]
Let $C_{\eta\eta}=Q\Lambda_\lambda Q^{\prime}$ be the eigendecomposition of $C_{\eta\eta}$. A
particular choice for $\tilde{A}$ is 
$\tilde{A}_{\ast}=C_{\eta\eta}^{-1/2}=Q\Lambda_\lambda^{-1/2}Q^{\prime}$,
which is symmetric and satisfies the requirement $\tilde{A}_{\ast}^{\prime}C_{\eta\eta}\tilde{A}_{\ast}=I_n$, and
it follows that 
\[
\varrho(\tilde{A}_{\ast})=\operatorname{tr}\{Q\Lambda_\lambda^{-1/2}Q^{\prime}Q\Lambda_\lambda Q^{\prime}\}=\operatorname{tr}\{\Lambda_\lambda^{1/2}\}=\sum_{i=1}^{n}\lambda_{i}^{1/2}.
\]
The corresponding vector of structural shocks is given by $u^{\ast}=\tilde{A}_{\ast}^{\prime}\eta$.

Let $\tilde{A}$ be an arbitrary matrix that satisfies $\tilde{A}^{\prime}C_{\eta\eta}\tilde{A}=I_n$, such that $\operatorname{var}(u)=I_n$ for $u=\tilde{A}^{\prime}\eta$. Because $\tilde{A}$ and $\tilde{A}_{\ast}$ are both full-rank
matrices, there exists a unique $R\in\mathbb{R}^{n\times n}$ for which $\tilde{A}=\tilde{A}_{\ast}R$, and it follows that 
$\operatorname{var}(u)=R^{\prime}\operatorname{var}(u^{\ast})R$.
Moreover, since $\operatorname{var}(u)=\operatorname{var}(u^\ast)=I_n$ it follows that $RR^{\prime}=I_n$, so that $R$ is an orthonormal matrix, and we will sometimes refer to $R$ as a rotation matrix. 
Next, define $M=Q^{\prime}R^{\prime}Q$,
which is a product of orthonormal matrices, such that $M$ is also
an orthonormal matrix (this can also be verified directly with $MM^{\prime}=Q^{\prime}R^{\prime}QQ^{\prime}RQ=I_n$).
We now have the identity
\begin{align*}
\varrho(\tilde{A}) = \operatorname{tr}\{\tilde{A}^{\prime}C_{\eta\eta}\} &= \operatorname{tr}\{R^{\prime}Q\Lambda_\lambda^{-1/2}Q^{\prime}Q\Lambda_\lambda Q^{\prime}\} \\ 
&=  \operatorname{tr}\{Q^{\prime}R^{\prime}Q\Lambda_\lambda^{1/2}\} =\sum_{i=1}^{n}M_{ii}\lambda_{i}^{1/2} 
\leq \sum_{i=1}^{n}\lambda_{i}^{1/2} = \varrho(\tilde{A}_{\ast}).
\end{align*}
The inequality follows from the fact that an orthonormal matrix
satisfies $\max_{i,j}|M_{ij}|\leq1$. This proves that $\tilde{A}_{\ast}$
leads to the maximal average correlation between the elements of $u^{\ast}=\tilde{A}_{\ast}^\prime\eta$
and the corresponding elements of $\eta$.

To complete the proof, we observe that $\operatorname{corr}(u,\varepsilon)=\operatorname{cov}(u,\varepsilon)\Lambda_{\sigma_\varepsilon}^{-1}=\operatorname{cov}(u,\Lambda_{\sigma_\varepsilon}^{-1}\varepsilon)=\operatorname{cov}(u,\eta)=\operatorname{corr}(u,\eta)$,
where $\eta=\Lambda_{\sigma_\varepsilon}^{-1}\varepsilon$. Hence $A_{\ast}=\Lambda_{\sigma_\varepsilon}^{-1}\tilde{A}_{\ast}=\Lambda_{\sigma_\varepsilon}^{-1}Q\Lambda_\lambda^{-1/2}Q^{\prime}$
maximizes the average correlation between $u_{i}$ and $\varepsilon_{i}$.
\hfill$\square$\medskip{} 

\noindent {\bf Proof of Proposition \ref{prop:IRF}.} 
The impulse response function under a general identification scheme $A \in \mathcal{A}$ is defined by $\operatorname{IRF}(h) = \Psi_h (A^\prime)^{-1}$, such that it is $\operatorname{IRF}^\ast(h) = \Psi_h (A_\ast^\prime)^{-1}$ for OASIS.

The right-hand side of the stated expression equals,
$$ \operatorname{IRF}^\ast(h) R = \Psi_h (A_\ast^\prime)^{-1} (A_\ast^{-1} A) = \Psi_h (A_\ast A_\ast^\prime)^{-1} A =  \Psi_h \Sigma_{\varepsilon\varepsilon} A, 
$$
where we used $(A_\ast A_\ast^\prime)^{-1} = \Sigma_{\varepsilon\varepsilon}$ because $A_\ast^\prime \Sigma_{\varepsilon\varepsilon} A_\ast = I$. For the same reason, we have $\Sigma_{\varepsilon\varepsilon} = (A^\prime)^{-1} A^{-1}$, such that 
$$ 
    \operatorname{IRF}^\ast(h) R = \Psi_h \left((A^\prime)^{-1} A^{-1}\right) A = \Psi_h (A^\prime)^{-1} = \operatorname{IRF}(h). 
$$
Finally,
$$ 
    R^\prime R 
    = (A_\ast^{-1} A)^\prime (A_\ast^{-1} A) 
    = A^\prime (A_\ast A_\ast^\prime)^{-1} A 
    = A^\prime \Sigma_{\varepsilon\varepsilon} A 
    = I
$$
shows that $R$ is indeed an orthonormal rotation matrix. This completes the proof.
\hfill$\square$\medskip{}

\noindent {\bf Proof of Theorem \ref{thm:Proximity}.} 
Let $\epsilon_{1},\ldots,\epsilon_{n}$ be the eigenvalues of $E=C_{\varepsilon\varepsilon}-I_n$.
Then $\sum_{i=1}^{n}\epsilon_{i}=\operatorname{tr}(E)=0$ and $d(C_{\varepsilon\varepsilon})=\frac{1}{n}\|E\|_{F}^{2}=\frac{1}{n}\sum_{i=1}^{n}\epsilon_{i}^{2}$,
where the last identity uses that $E$ is symmetric. Moreover, the
eigenvalues of $C_{\varepsilon\varepsilon}$ are related to those of $E$, by $\lambda_{i}=1+\epsilon_{i}$,
for $i=1,\ldots,n$. For OASIS, $\tilde{A}_\ast=Q\,\Lambda_\lambda^{-1/2}Q^{\prime}$, we have by Corollary~\ref{cor:OASIS} that 
$\operatorname{tr}(\tilde{A}_\ast^\prime C_{\varepsilon\varepsilon})=\sum_{i=1}^{n}\lambda_{i}^{1/2}.$ The Taylor series for an element in this sum is:
\begin{equation}
\lambda_{i}^{1/2}=(1+\epsilon_{i})^{1/2}=1+\tfrac{1}{2}\epsilon_{i}-\tfrac{1}{8}\epsilon_{i}^{2}+\tfrac{1}{16}\epsilon_{i}^{3}-\cdots,\label{eq:TaylorSqrt}
\end{equation}
and by adding these up, we find
$$
\bar{\rho}_\ast=\tfrac{1}{n}\operatorname{tr}(\tilde{A}_\ast^\prime C_{\varepsilon\varepsilon})  =  1-\tfrac{1}{8n}\sum_{i=1}^{n}\epsilon_{i}^{2}+O\left(\tfrac{1}{n}\sum_{i=1}^{n}\epsilon_{i}^{3}\right)\\
  =  1-\tfrac{1}{8}d(C_{\varepsilon\varepsilon})+\tfrac{1}{n}O(\operatorname{tr}\{E^{3}\}),
$$
where we used $\sum_{i=1}^{n}\epsilon_{i}^{3}=\operatorname{tr}(E^{3})$. 

Next, we consider $\tilde{A}_\mathrm{c}^\prime=L^{-1}$ based on the Cholesky factorization,
$C_{\varepsilon\varepsilon}=L\,L^{\prime}$, where $L$ is lower triangular (with positive
diagonal entries). It follows that $\tilde{A}_\mathrm{c}^\prime C_{\varepsilon\varepsilon}=L^{-1}L\,L^{\prime}=L^{\prime}$,
so that 
$
\operatorname{tr}(\tilde{A}_\mathrm{c}^\prime C_{\varepsilon\varepsilon})=\operatorname{tr}(L)=\sum_{i=1}^{n}L_{ii}.$ 
From the identity, $1=[C_{\varepsilon\varepsilon}]_{ii}=\sum_{j\leq i}L_{ij}^{2}$, we have 
\begin{equation}
L_{ii}=\sqrt{1-\sum_{j<i}L_{ij}^{2}}=1-\tfrac{1}{2}\sum_{j<i}L_{ij}^{2}-\tfrac{1}{8}\left(\sum_{j<i}L_{ij}^{2}\right)^{2}-\cdots,\label{eq:Lii}
\end{equation}
where we used the same Taylor expansion as in (\ref{eq:TaylorSqrt}). 
Next, $L_{ij}=[C_{\varepsilon\varepsilon}]_{ij}+O(E^{2})=E_{ij}+O(E^{2})$, such that summing
over $i$ in (\ref{eq:Lii}) yields 
\[
\operatorname{tr}(L)=n-\tfrac{1}{2}\sum_{i=1}^{n}\sum_{j<i}E_{ij}^{2}+O\left(\operatorname{tr}\{E^{3}\}\right).
\]
Since $d(C_{\varepsilon\varepsilon})=\tfrac{2}{n}\sum_{i>j}E_{ij}^{2}$, we can conclude
that 
\[
\tfrac{1}{n}\operatorname{tr}(\tilde{A}_\mathrm{c}^\prime C_{\varepsilon\varepsilon})=1-\tfrac{1}{4}d(C_{\varepsilon\varepsilon})+O\left(\operatorname{tr}\{E^{3}\}\right).
\]
Combined, we have 
\[
\frac{1-\bar\rho_{\mathrm{c}}}{1-\bar\rho_\ast}
=2
\frac{1+O\left(\operatorname{tr}\{E^3\}/d(C_{\varepsilon\varepsilon})\right)}
     {1+O\left(\operatorname{tr}\{E^3\}/d(C_{\varepsilon\varepsilon})\right)}
=2\left(1+O\left(\|E\|_F\right)\right)
=2\left(1+O\left(\sqrt{d(C_{\varepsilon\varepsilon})}\right)\right),
\]
using $\operatorname{tr}\{E^3\}=O(\|E\|_F^3)$ and $d(C_{\varepsilon\varepsilon})=\|E\|_F^2/n=O(\|E\|_F^2)$. 
\hfill$\square$ \medskip{}

\noindent {\bf Proof of Corollary \ref{cor:Equicorr}.} 
From  \citet{ArchakovHansen:CanonicalBlockMatrix} we have that the $r$-th power of an equicorrelation matrix can be expressed as
\begin{equation}
C^{r}=(1+(n-1)\rho)^{r}P_{n}+(1-\rho)^{r}P_{n}^\perp.\label{eq:Cpower}
\end{equation}
where $P_{n}=\iota_n(\iota_n^{\prime}\iota_n)^{-1}\iota_n^{\prime}=\tfrac{1}{n}\iota_n \iota_n^\prime$ and $P_{n}^\perp=I-P_{n}$ are orthogonal projection matrices and $\iota_n\in\mathbb{R}^n$ is the vector of ones. So, 
(\ref{eq:EquiCorrOasis}) follows directly from (\ref{eq:Cpower}) with $r=1/2$,
\[
\bar{\rho}_\ast=\tfrac{1}{n}\operatorname{tr}\{C_{\varepsilon\varepsilon}^{1/2}\}=\tfrac{1}{n}\sqrt{1+(n-1)\rho}+\sqrt{1-\rho}\left(1-\tfrac{1}{n}\right).
\]
For the Cholesky decomposition, $C_{\varepsilon\varepsilon}=LL^{\prime}$, we have $\bar{\rho}_{\mathrm{c}}=\frac{1}{n}\operatorname{tr}\{L\}=\frac{1}{n}\sum_{k=1}^{n}L_{kk}$.
A diagonal element $L_{kk}$ is the standard deviation of $\eta_{k}$
after controlling for $\eta_{1},\ldots,\eta_{k-1}$, where $\operatorname{var}(\eta)=C_{\eta\eta}$.
Standard projection arguments give us the linear regression, $\eta_{k}=\rho\iota_{k-1}^{\prime}\tilde{C}_{[k-1]}^{-1}\eta_{1:k-1}+L_{kk}u_{k}$,
where $\tilde{C}_{[k-1]}=\operatorname{var}(\eta_{1:k-1})$ with $\eta_{1:k-1}=(\eta_{1},\ldots,\eta_{k-1})^{\prime}$, and
the error variance is given by
\begin{align*}
L_{kk}^{2} & =1-\rho^{2}\iota_{k-1}\tilde{C}_{[k-1]}^{-1}\iota_{k-1}^{\prime}\\
 & =1-\rho^{2}\iota_{k-1}\left[(1+(k-2)\rho)^{-1}P_{k-1}+(1-\rho)^{-1}\left(P_{k-1}^\perp\right)\right]\iota_{k-1}^{\prime}\\
 & =1-\rho^{2}(k-1)/(1+(k-2)\rho),
\end{align*}
where we used (\ref{eq:Cpower}) with $r=-1$. This proves (\ref{eq:EquiCorrChol}). 
\hfill$\square$ \medskip{}

\noindent {\bf Proof of Corollary \ref{cor:CholeskySimilar}.}
Follows directly from Theorem \ref{thm:Proximity}.
\hfill$\square$ \medskip{}

\noindent {\bf Proof of Theorem \ref{thm:ProxyVAR}.} 
Let $\tilde{\boldsymbol{a}}\equiv C_{\varepsilon\varepsilon}^{1/2}\Lambda_{\sigma_\varepsilon} \boldsymbol{a}$ so that $\tilde{\boldsymbol{a}}^\prime \tilde{\boldsymbol{a}}=I_r$. Then
$$
 \operatorname{corr}(\boldsymbol{a}^\prime\varepsilon,z)\Lambda_w 
 = \boldsymbol{a}^\prime \Lambda_{\sigma_\varepsilon} C_{\varepsilon z}\Lambda_w
 = \boldsymbol{a}^\prime \Lambda_{\sigma_\varepsilon} C_{\varepsilon \varepsilon}^{1/2} C_{\varepsilon \varepsilon}^{-1/2}C_{\varepsilon z}\Lambda_w
 = \tilde{\boldsymbol{a}}^\prime \Xi ,    
$$
such that 
$$
g(\boldsymbol{a}_\ast) 
    = \operatorname{tr}\{\boldsymbol{a}_\ast^\prime \Lambda_{\sigma_\varepsilon} C_{\varepsilon\varepsilon}^{1/2} \Xi \}
    = \operatorname{tr}\{VU^\prime U \Lambda_\xi V^\prime \}=\sum_{i=1}^r\xi_i,
$$
and 
$$
g(\boldsymbol{a}) 
    = \operatorname{tr}\{\tilde{\boldsymbol{a}}^\prime \Xi \}
    =\operatorname{tr}\{\tilde{\boldsymbol{a}}^\prime U\Lambda_\xi V^\prime \} 
    = \operatorname{tr}\{V^\prime \tilde{\boldsymbol{a}}^\prime U\Lambda_\xi \} 
    = \sum_{i=1}^rM_{ii}\xi_i,
$$
where $M_{ii}$, $i=1,\ldots,r$ are the diagonal elements of $M\equiv V^\prime \tilde{\boldsymbol{a}}^\prime U$. Thus, $M_{ii}=v_i^{\prime}  \tilde{\boldsymbol{a}}^\prime u_i$, where $v_i$ and $u_i$ are the $i$-th columns of $V$ and $U$, respectively. Because $\tilde{\boldsymbol a}^\prime\tilde{\boldsymbol a}=I_r$ and $\|v_i\|=1$ we have 
$
\|\tilde{\boldsymbol a}v_i\|^2
= v_i'(\tilde{\boldsymbol a}'\tilde{\boldsymbol a})v_i
= v_i'v_i
= 1,
$
such that
$$
|M_{ii}| = |v_i'\tilde{\boldsymbol a}'u_i | 
    \leq  \|u_i\|\,\|\tilde{\boldsymbol a}v_i\| = 1,
$$ by the Cauchy-Schwarz inequality and $\|u_i\|=1$. 
Because $\xi_i\geq 0$ for all $i$ (by the definition of the SVD) we have shown that
$g(\boldsymbol{a}) \leq g(\boldsymbol{a_\ast})$. 

Moreover, if $C_{\varepsilon z}$ has full column rank, then $\xi_i>0$ for all $i$, and 
$$
g(\boldsymbol{a}) = g(\boldsymbol{a_\ast})\Leftrightarrow M_{ii}=1,\ \forall i \Leftrightarrow u_i=\tilde{\boldsymbol{a}}v_i,\ \forall i\Leftrightarrow U=\tilde{\boldsymbol{a}}V,
$$ 
which implies $\tilde{\boldsymbol{a}}=UV^\prime=\tilde{\boldsymbol{a}}_\ast$. This shows that $\boldsymbol{a}_\ast$ is the unique maximizer in this case.
\hfill$\square$ \medskip{}

\noindent{}{\bf Proof of Corollary \ref{cor:ProxyVarFOC}} It is straightforward to verify $(i)$. For $(ii)$ we first observe that
$\Lambda_{\sigma_\varepsilon} \boldsymbol{a}_\ast \boldsymbol{a}_\ast^\prime \Lambda_{\sigma_\varepsilon}=C_{\varepsilon\varepsilon}^{-1/2}UU^\prime C_{\varepsilon\varepsilon}^{-1/2}$. Multiply the left-hand side of (\ref{eq:GMM2pop}) by $C_{\varepsilon\varepsilon}^{-1/2}$ from left and by $\Lambda_w$ from right (both full rank matrices) to get
$$C_{\varepsilon\varepsilon}^{-1/2}C_{\varepsilon z}\Lambda_w-UU^\prime C_{\varepsilon\varepsilon}^{-1/2}C_{\varepsilon z}\Lambda_w=\Xi-UU^\prime \Xi=\Xi-\Xi=0,$$
where we used $UU^\prime\Xi=UU^\prime U \Lambda_\xi V^\prime=\Xi$. This proves (\ref{eq:GMM2pop}). Finally, $(iii)$ follows from $$\operatorname{corr}(u^\ast_t,z_t)\Lambda_w=\boldsymbol{a}_\ast^\prime\Lambda_{\sigma_\varepsilon}\operatorname{corr}(\varepsilon_t,z_t)\Lambda_w
=VU^\prime C_{\varepsilon\varepsilon}^{-1/2}\Lambda_{\sigma_\varepsilon}^{-1}\Lambda_{\sigma_\varepsilon} C_{\varepsilon z}\Lambda_w
=VU^\prime\Xi
=V\Lambda_\xi V^\prime.$$
\hfill$\square$ \medskip{}

\noindent{}{\bf Proof of Lemma \ref{lem:LinProject}.}
From standard linear projection results, it follows directly that $\eta$ has mean zero and $\mathbb{E}[u\eta^\prime]=0$. So, the last statement is the stronger condition: $\mathbb{E}[\varepsilon \eta^\prime]=0$. 
Let $(u^\prime,v^\prime)^\prime=[\boldsymbol{a}_0,\boldsymbol{b}_0]^\prime \varepsilon = A_0^\prime \varepsilon$ with $A_0^\prime \Sigma_{\varepsilon\varepsilon} A_0 = I_n$. Then $A_0^{-1}=A_0^\prime \Sigma_{\varepsilon\varepsilon}$ such that $\varepsilon = \Sigma_{\varepsilon\varepsilon} \boldsymbol{a}_0 u + \Sigma_{\varepsilon\varepsilon} \boldsymbol{b}_0 v$ 
and
$\Sigma_{z\varepsilon} =  \Sigma_{zu} \boldsymbol{a}_0^\prime \Sigma_{\varepsilon\varepsilon} + \Sigma_{zv} \boldsymbol{b}_0^\prime \Sigma_{\varepsilon\varepsilon} = \Phi \boldsymbol{a}_0^\prime \Sigma_{\varepsilon\varepsilon}$, where we used Assumption 2. It follows that $\eta = z - \mu_z - \Phi u = z - \mu_z - \Sigma_{z\varepsilon} \Sigma_{\varepsilon\varepsilon}^{-1} \varepsilon$, so $\mathbb{E}[\eta] = 0$ and $\mathbb{E}[\varepsilon\eta^\prime] = 0$ by linear projection orthogonality.
\hfill$\square$ \medskip{}

\noindent{}{\bf Proof of Theorem \ref{thm:OASISrecoversRotation}.} 
From the identity, 
$\Sigma_{z\varepsilon } 
    =  \Phi\boldsymbol{a}_0^\prime\Sigma_{\varepsilon\varepsilon}
$, we have that $\operatorname{corr}(\varepsilon,z)$ equals
\begin{equation}
    C_{\varepsilon z} =     \Lambda_{\sigma_\varepsilon}^{-1}\Sigma_{\varepsilon z}
    \Lambda_{\sigma_z}^{-1}
    =
    C_{\varepsilon\varepsilon} \Lambda_{\sigma_\varepsilon} \boldsymbol{a}_0 \Phi^\prime \Lambda_{\sigma_z}^{-1},
\end{equation}
and that
$$
\Xi
= C_{\varepsilon\varepsilon}^{-1/2} C_{\varepsilon z} \Lambda_w
= C_{\varepsilon\varepsilon}^{-1/2} (C_{\varepsilon\varepsilon} \Lambda_{\sigma_\varepsilon} \boldsymbol{a}_0 \Phi^\prime \Lambda_{\sigma_z}^{-1}) \Lambda_w 
    = C_{\varepsilon\varepsilon}^{1/2} \Lambda_{\sigma_\varepsilon} \boldsymbol{a}_0 \Phi^\prime \Lambda_{\sigma_z}^{-1} \Lambda_w
    = C_{\varepsilon\varepsilon}^{1/2} \Lambda_{\sigma_\varepsilon} \boldsymbol{a}_0 M.
$$
Because $M$ is symmetric and positive definite, we have $M=\Phi^\prime\Lambda_{\sigma_z}^{-1} \Lambda_w=Q\Lambda_\psi Q^\prime$ where $Q^\prime Q=I_r$ and $\Lambda_\psi$ is the diagonal matrix with positive eigenvalues. 

Next, set $W = C_{\varepsilon\varepsilon}^{1/2} \Lambda_{\sigma_\varepsilon} \boldsymbol{a}_0 Q$ and observe that
\begin{align*}
    W^\prime W &= Q^\prime \boldsymbol{a}_0^\prime \Lambda_{\sigma_\varepsilon} C_{\varepsilon\varepsilon}^{1/2} C_{\varepsilon\varepsilon}^{1/2} \Lambda_{\sigma_\varepsilon} \boldsymbol{a}_0 Q \\
    &= Q^\prime \boldsymbol{a}_0^\prime (\Lambda_{\sigma_\varepsilon} C_{\varepsilon\varepsilon} \Lambda_{\sigma_\varepsilon}) \boldsymbol{a}_0 Q \\
    &= Q^\prime \boldsymbol{a}_0^\prime \Sigma_{\varepsilon\varepsilon} \boldsymbol{a}_0 Q = Q^\prime I_r Q  = I_r.
\end{align*}
So it follows that $\Xi = U\Lambda_\xi V^\prime$ with $U = W$, $\Lambda_\xi = \Lambda_\psi$, and $V = Q$.

Finally, from Theorem \ref{thm:ProxyVAR} we have
$$
    \boldsymbol{a}_\ast = \Lambda_{\sigma_\varepsilon}^{-1} C_{\varepsilon\varepsilon}^{-1/2} U V^\prime = \Lambda_{\sigma_\varepsilon}^{-1} C_{\varepsilon\varepsilon}^{-1/2} W Q^\prime 
    = \Lambda_{\sigma_\varepsilon}^{-1} C_{\varepsilon\varepsilon}^{-1/2} C_{\varepsilon\varepsilon}^{1/2} \Lambda_{\sigma_\varepsilon} \boldsymbol{a}_0 Q Q^\prime = \boldsymbol{a}_0.
$$
Finally, if $\Phi$ is diagonal, we have $\Lambda_\xi=\Lambda_w\Lambda_{\sigma_z}^{-1} \Phi$ and $V=I_r$.
\hfill$\square$ \medskip{}

\noindent {\bf Proof of Lemma \ref{lem:ZeInv(ee)eZ}.}
From standard linear projection results, it follows directly that $\eta$ satisfies
$\mathbb{E}[\eta]=0$ and $\mathbb{E}[\eta\varepsilon^\prime]=0$.

Let $(u^\prime,v^\prime)^\prime = [\boldsymbol{a}_0,\boldsymbol{b}_0]^\prime \varepsilon = A_0^\prime \varepsilon$, where
$
A_0^\prime \Sigma_{\varepsilon\varepsilon} A_0 = I_n$.
Then
$A_0^{-1}=A_0^\prime \Sigma_{\varepsilon\varepsilon}$,
such that
$\varepsilon = \Sigma_{\varepsilon\varepsilon}\boldsymbol{a}_0 u
+ \Sigma_{\varepsilon\varepsilon}\boldsymbol{b}_0 v$. 
Therefore,
$$
\Sigma_{Z\varepsilon}
=
\operatorname{cov}(Z,\varepsilon)
=
\operatorname{cov}(Z,u)\boldsymbol{a}_0^\prime \Sigma_{\varepsilon\varepsilon}
+
\operatorname{cov}(Z,v)\boldsymbol{b}_0^\prime \Sigma_{\varepsilon\varepsilon}.
$$
The last term is zero by Assumption 2$^\prime$, which proves
$\Sigma_{Z\varepsilon} = \boldsymbol{\Phi}\boldsymbol{a}_0^\prime \Sigma_{\varepsilon\varepsilon}$.

Substituting this identity into the linear projection gives
$$
Z
=
\mu_Z+\Sigma_{Z\varepsilon}\Sigma_{\varepsilon\varepsilon}^{-1}\varepsilon+\eta
=
\mu_Z+\boldsymbol{\Phi}\boldsymbol{a}_0^\prime \varepsilon+\eta
=
\mu_Z+\boldsymbol{\Phi}u+\eta,
$$
which proves the claimed representation.

Finally,
$$
\Sigma_{Z\varepsilon}\Sigma_{\varepsilon\varepsilon}^{-1}\Sigma_{\varepsilon Z}
=
\boldsymbol{\Phi}\boldsymbol{a}_0^\prime \Sigma_{\varepsilon\varepsilon}
\Sigma_{\varepsilon\varepsilon}^{-1}
\Sigma_{\varepsilon\varepsilon}\boldsymbol{a}_0\boldsymbol{\Phi}^\prime
=
\boldsymbol{\Phi}\bigl(\boldsymbol{a}_0^\prime \Sigma_{\varepsilon\varepsilon}\boldsymbol{a}_0\bigr)\boldsymbol{\Phi}^\prime
=
\boldsymbol{\Phi}\boldsymbol{\Phi}^\prime,
$$
where we used $\boldsymbol{a}_0^\prime \Sigma_{\varepsilon\varepsilon}\boldsymbol{a}_0=I_r$. This completes the proof. $\square$\medskip{}

\noindent{}{\bf Proof of Lemma \ref{lem:Baseline}.}
Under Assumption 3 we have $\boldsymbol{\phi}_{ij} = \mathbf{0}$ for all $i \neq j$ such that
$$
    Z_i = \mu_{Z_i} + \boldsymbol{\phi}_{ii} u_i + \eta_i.
$$
By Assumption 2' (Exclusion Restriction), the measurement error $\eta_i$ for the $i$-th block is uncorrelated with the reduced-form innovations $\varepsilon$, yielding $\mathbb{E}[\varepsilon\eta_i^\prime] = \mathbf{0}$.

To derive the cross-covariance $\Sigma_{\varepsilon Z_i}$, we evaluate the expectation:
$$
    \Sigma_{\varepsilon Z_i} = \mathbb{E}[\varepsilon (Z_i - \mu_{Z_i})^\prime] = \mathbb{E}[\varepsilon (\boldsymbol{\phi}_{ii} u_i + \eta_i)^\prime] = \mathbb{E}[\varepsilon u_i]\boldsymbol{\phi}_{ii}^\prime + \mathbb{E}[\varepsilon \eta_i^\prime].
$$
Since $\mathbb{E}[\varepsilon \eta_i^\prime] = \mathbf{0}$, this reduces to $\mathbb{E}[\varepsilon u_i]\boldsymbol{\phi}_{ii}^\prime$.

Let $\boldsymbol{a}_{0,i}$ denote the true structural weighting vector for the $i$-th shock, such that $u_i = \boldsymbol{a}_{0,i}^\prime \varepsilon$. We can rewrite the expectation $\mathbb{E}[\varepsilon u_i]$ as
$
    \mathbb{E}[\varepsilon u_i] = \mathbb{E}[\varepsilon (\boldsymbol{a}_{0,i}^\prime \varepsilon)^\prime] = \mathbb{E}[\varepsilon \varepsilon^\prime \boldsymbol{a}_{0,i}] = \Sigma_{\varepsilon\varepsilon}\boldsymbol{a}_{0,i}
$.
Substituting this result back into the cross-covariance expression gives:
$$
    \Sigma_{\varepsilon Z_i} = \Sigma_{\varepsilon\varepsilon}\boldsymbol{a}_{0,i}\boldsymbol{\phi}_{ii}^\prime.
$$
This completes the proof.
\hfill$\square$ \medskip{}

\noindent{\bf Proof of Theorem \ref{thm:OptimalBeta}.}
Let $G_i=Q\Lambda Q^\prime $ be the eigendecomposition of $G_i$. It is a standard result for Rayleigh quotients that the principal eigenvector, $q_i^\ast$, maximizes $g_i(q)\equiv q^\prime G_i q/(q^\prime q)$ with maximum value $g_i(q^\ast_i)=\lambda^\ast_i=\alpha_i^2$, and satisfies $q^{\ast\prime}_i q^\ast_i=1$. Because $f_i(b) = g_i(\Sigma_{Z_i Z_i}^{1/2} b)$, it immediately follows that any vector of the form $b \propto \Sigma_{Z_i Z_i}^{-1/2} q_i^\ast$ maximizes $f_i$. Moreover, 
$\operatorname{var}(\bar{z}_i) = \beta_i^{\ast\prime} \Sigma_{Z_i Z_i} \beta_i^\ast = (\tau_i q_i^{\ast\prime} \Sigma_{Z_i Z_i}^{-1/2}) \Sigma_{Z_i Z_i} (\tau_i \Sigma_{Z_i Z_i}^{-1/2} q_i^\ast) = \tau_i^2 q_i^{\ast\prime} q_i^\ast = 1,
$ verifies the unit variance. 

To resolve the sign indeterminacy, we use that
$\Sigma_{Z_i \varepsilon} \Sigma_{\varepsilon \varepsilon}^{-1} \Sigma_{\varepsilon Z_i} = \boldsymbol{\phi}_{ii} \boldsymbol{\phi}_{ii}^\prime$ has rank one, and multiplying by $\Sigma_{Z_i Z_i}^{-1/2}$ on both sides yields $G_i = \nu_i \nu_i^\prime$, where $\nu_i = \Sigma_{Z_i Z_i}^{-1/2} \boldsymbol{\phi}_{ii}$ such that 
$\boldsymbol{\phi}_{ii} = \pm \alpha_i \Sigma_{Z_i Z_i}^{1/2} q_i^\ast$. The first element of $\boldsymbol{\phi}_{ii}$ is strictly positive: $[\boldsymbol{\phi}_{ii}]_1 > 0$, by Assumption 1$^\prime$ (Anchor Proxy). This is indeed satisfied by setting 
$\boldsymbol{\phi}_{ii} = \tau_i \alpha_i \Sigma_{Z_i Z_i}^{1/2} q_i^\ast$, and 
$$
    \operatorname{cov}(Z_i, \bar{z}_i) = \Sigma_{Z_i Z_i} \beta_i^\ast = \tau_i \Sigma_{Z_i Z_i}^{1/2} q_i^\ast = \alpha_i^{-1} \boldsymbol{\phi}_{ii},
$$
confirms that the directional sign of the optimal linear combination is consistent $[\boldsymbol{\phi}_{ii}]_1 > 0$ and guarantees $\bar{\Phi}_{ii}=\operatorname{corr}(u_i, \beta_i^{\ast\prime} Z_i) = \operatorname{corr}(u_i, \bar{z}_i)> 0$.  Since, $\bar{\Phi}_{ij}=0$ for $i\neq j$, it follows that $\bar{\Phi}$ is a positive definite diagonal matrix, and the conditions of Theorem \ref{thm:OASISrecoversRotation} are satisfied with any choice of positive weights. 
\hfill$\square$ \medskip{}

\noindent{}{\bf Proof of Lemma \ref{lem:ZieInv(ee)eZi}.}
By Lemma \ref{lem:ZeInv(ee)eZ} and Assumption 1'' we have $
    \Sigma_{Z_i\varepsilon}\Sigma_{\varepsilon\varepsilon}^{-1}\Sigma_{\varepsilon Z_i}
    = \sum_{j=1}^r \boldsymbol{\phi}_{ij}\boldsymbol{\phi}_{ij}^\prime
    = \boldsymbol{\phi}_{ii}\boldsymbol{\phi}_{ii}^\prime \sum_{j=1}^r s_{ij}^2
    = s_{i\bullet}^2 \boldsymbol{\phi}_{ii}\boldsymbol{\phi}_{ii}^\prime.
    $
\hfill$\square$ \medskip{}

\begin{lemma}\label{lem:Convex}
Let  $H$ be an $n\times n$ symmetric positive definite, let $\iota\in\mathbb{R}^n$ be the vector of ones, and let $\Lambda_x=\operatorname{diag}(x_1,\ldots,x_n)$. Then   
\begin{equation}\label{eq:AppendixFOC}
     \operatorname{diag}( [\Lambda_xH \Lambda_x]^{1/2}) = \iota, \qquad \text{for } x>0
\end{equation}
has a unique positive solution, $x_1^\ast,\ldots x_n^\ast$. 
\end{lemma}
\begin{proof}
Consider the scalar-valued objective function,
$$f(y) = \operatorname{tr}(G(y)) - \iota^\prime y,\qquad G(y) \equiv \left(e^{\Lambda_y} H e^{\Lambda_y}\right)^{1/2}.$$
The proof amounts to showing that $f : \mathbb{R}^n \to \mathbb{R}$ 
possesses a unique global minimum $y^\ast \in \mathbb{R}^n$, whose first-order condition coincides exactly with (\ref{eq:AppendixFOC}) for  $x_i = \exp(y_i)$ for $i=1,\ldots,n$.

We seek the gradient of $f(y)$. The differential of $G^2 = e^{\Lambda_y} H e^{\Lambda_y}$ with respect to the diagonal matrix $\Lambda_y$ is:
$$d(G^2) = (d\Lambda_y) e^{\Lambda_y} H e^{\Lambda_y} + e^{\Lambda_y} H e^{\Lambda_y} (d\Lambda_y) = (d\Lambda_y) G^2 + G^2 (d\Lambda_y)$$
By the product rule, we also have $d(G^2) = G(dG) + (dG)G$. Equating these expressions gives
\begin{equation}\label{eq:Sylvester}
    G(dG) + (dG)G = (d\Lambda_y) G^2 + G^2 (d\Lambda_y).
\end{equation}
Next,  multiplying by $G^{-1}$ from left and taking the trace yields:
$\operatorname{tr}(dG) + \operatorname{tr}(G^{-1}(dG)G)  = \operatorname{tr}(G^{-1} d\Lambda_y G^2) + \operatorname{tr}(G d\Lambda_y)$, such that 
$$d\operatorname{tr}(G) = \operatorname{tr}(G d\Lambda_y) = \sum_{i=1}^n G_{ii} dy_i.$$
This shows that the gradient of $f(y)$ is $\nabla f(y) = \operatorname{diag}(G(y)) - \iota$, such that the FOC of this optimization problem satisfies (\ref{eq:AppendixFOC}).

The uniqueness now follows by showing that $f(y)$ is strictly convex. The quadratic form of the Hessian is given by the differential of the gradient:
$$dy^\prime \nabla^2 f(y) dy = \sum_{i=1}^n dy_i dG_{ii} = \operatorname{tr}((d\Lambda_y)(dG))$$
Let $G = Q \Lambda_\kappa Q^\prime$ be the eigendecomposition of $G$, where $\Lambda_\kappa = \operatorname{diag}(\kappa_1,\ldots,\kappa_n)$ contains strictly positive eigenvalues. Defining $\widetilde{dG} = Q^\prime dG Q$ and $\widetilde{d\Lambda_y} = Q^\prime d\Lambda_y Q$, we pre-multiply (\ref{eq:Sylvester}) by $Q^\prime$ and post-multiply by $Q$ to obtain:
$$
\Lambda_\kappa \widetilde{dG} + \widetilde{dG} \Lambda_\kappa = \widetilde{d\Lambda_y} \Lambda_\kappa^2 + \Lambda_\kappa^2 \widetilde{d\Lambda_y}.
$$
Solving for the elements of $\widetilde{dG}$ gives:
$$
[\widetilde{dG}]_{kl} = [\widetilde{d\Lambda_y}]_{kl} \frac{\kappa_k^2 + \kappa_l^2}{\kappa_k + \kappa_l}
$$
Substituting this into the trace formulation:
$$\operatorname{tr}((d\Lambda_y)(dG)) = \operatorname{tr}(\widetilde{d\Lambda_y} \widetilde{dG}) = \sum_{k=1}^n \sum_{l=1}^n ([\widetilde{d\Lambda_y}]_{kl})^2 \frac{\kappa_k^2 + \kappa_l^2}{\kappa_k + \kappa_l}$$
Because $\kappa_k > 0$ for all $k$, this sum is strictly positive for any non-zero differential $dy$. Therefore, the Hessian $\nabla^2 f(y)$ is strictly positive definite everywhere. Since $f(y)$ is strictly convex on $\mathbb{R}^n$, it has at most one stationary point $y^\ast$. 

The existence of a minimum for $f$ follows from the fact that
$f(y)\ge \sqrt{m}\sum_{i=1}^n e^{y_i}-\sum_{i=1}^n y_i$, where $m=\lambda_{\min}(H)>0$, such that the right-hand side tends to $+\infty$ whenever any coordinate $y_i\to\pm\infty$. Since $f$ is continuous, it attains a global minimum at some $y^*\in\mathbb{R}^n$.

Consequently, the transformation $x_i^\ast = \exp(y_i^\ast)$ yields the unique positive solution to the scaling equation.
\end{proof}
\noindent{\bf Proof of Theorem \ref{thm:ProxyOasisLeakage}.}
First, the elements of the composite signal matrix $\bar{\Phi}$ are given by:
$$ 
    \operatorname{cov}(\bar{z}_i,u_j) = \beta_i^{\ast\prime} \operatorname{cov}(Z_i, u_j) = \beta_i^{\ast\prime} \boldsymbol{\phi}_{ij}= s_{ij}(\beta_i^{\ast\prime}\boldsymbol{\phi}_{ii})= s_{ij}\alpha_i,
$$
where we used Assumption 1'' (Proportional Leakage) and (\ref{eq:betaStar}) ($\beta_i^{\ast\prime}\boldsymbol{\phi}_{ii} = \alpha_i$). This proves $\bar{\Phi} = \Lambda_\alpha S$.

Second,  we can isolate $\Lambda_\alpha$ from the observable data as follows. Observe that,
$$
    H = \Sigma_{\bar{z}\varepsilon}\Sigma_{\varepsilon\varepsilon}^{-1}\Sigma_{\varepsilon\bar{z}} = \bar{\Phi} \bar{\Phi}^\prime
    = (\Lambda_\alpha S)(\Lambda_\alpha S)^\prime = \Lambda_\alpha S^2 \Lambda_\alpha,
$$ 
where we used the symmetry of $S$. Define,
$$
    \tilde{H}(x) \equiv \Lambda_x H \Lambda_x = (\Lambda_x \Lambda_\alpha) S^2 (\Lambda_\alpha \Lambda_x),
$$
and let $\tilde{H}^{1/2}(x)$ denote its symmetric square root. Then $\tilde{H}^{1/2}(x)= S$ if and only if $\Lambda_x = \Lambda_\alpha^{-1}$, and since $\operatorname{diag}(S)=\iota$, we have $\operatorname{diag}(\tilde{H}^{1/2}(x) ) = \iota $ if and only if $\Lambda_x = \Lambda_\alpha^{-1}$, where the uniqueness follows from Lemma \ref{lem:Convex}. 

Finally, with
 $\Lambda_w = \Lambda_\alpha^{-1}$, then $M=\Lambda_w\bar{\Phi}=S$ is symmetric and positive definite, and the conditions of Theorem \ref{thm:OASISrecoversRotation} are satisfied, such that OASIS  recovers the structural rotation $\boldsymbol{a}_\ast=\boldsymbol{a}_0$.
\hfill$\square$ \medskip{}

\newpage
\setcounter{table}{0}
\global\long\def\thetable{B.\arabic{table}}%
\section{Data Sources and Proxy Definitions}\label{app:data}

We used the original data for most of the studies revisited in our empirical analysis. The macroeconomic time series and reduced-form covariance matrices were extracted directly from the authors' replication files, which are compiled and available in our GitHub repository (\url{https://github.com/reinhardhansen/OASIS-Replication}). This appendix provides further details on the data utilized throughout the paper. Section B.1 outlines the specific sample periods for the standard SVAR studies, and Section B.2 details the definitions and sources of the proxy variables in \citet{StockWatson:2012}. 

\subsection{Sample Periods in SVAR Studies}

In four of the studies, we extended the analysis using larger sample periods to incorporate more recent macroeconomic data, as detailed in Table \ref{tab:sample_periods}.

\begin{table}[htbp!]
\caption{Sample Periods of Structural VAR Studies by Shock Type}
\label{tab:sample_periods}
{\setstretch{1.00}\footnotesize
    \begin{tabularx}{\textwidth}{@{}l X X @{}}
        \toprule
        \textbf{Study ID} & \textbf{Original Sample Period} & \textbf{Estimation Sample Period} \\
        \midrule
        \multicolumn{3}{c}{\textbf{Monetary Shocks}} \\
        B86 & 1953:Q1--1984:Q4 & 1959:Q1--2024:Q4 \\
        S95 & 1959:M1--1992:M2 & same \\
        LSZ96 & 1960:M1--1996:M3 & 1959:Q1--2018:Q2 \\
        CEE99 & 1965:Q3--1995:Q2 & same \\
        CEE05 & 1965:Q3--1995:Q3 & same \\
        BL09 & 1983:M1--2002:M12 & same \\[2mm]
        
        \multicolumn{3}{c}{\textbf{Fiscal Shocks}} \\
        BP02 & 1960:Q1--1994:Q4 & 1947:Q1--2018:Q3 \\
        RZ11 & 1954:Q4--2006:Q4 & same \\
        FG16 & 1981:Q3--2013:Q3 & same \\[2mm]
        
        \multicolumn{3}{c}{\textbf{Uncertainty Shocks}} \\
        B09 & 1962:M6--2008:M6 & same \\
        CCG14 & 1962:Q3--2012:Q3 & same \\
        BB17 & 1986:Q1--2014:Q4 & same \\
        BO23 & 1960:Q2--2018:Q2 & same \\[2mm]
        
        \multicolumn{3}{c}{\textbf{Financial Shocks}} \\
        GZ12 & 1973:Q3--2010:Q3 & same \\
        FGMS24 & 1973:M1--2019:M12 & same \\[2mm]
        
        \multicolumn{3}{c}{\textbf{Oil Price Shocks}} \\
        LS04 & 1972:Q1--2000:Q4 & same \\
        LP18 & 1976:M1--2014:M12 & same \\[2mm]
        
        \multicolumn{3}{c}{\textbf{Sectoral}} \\
        FGKV25 & 1988:Q1--2019:Q4 & same \\[2mm]
        
        \multicolumn{3}{c}{\textbf{Highly Correlated VARs}} \\
        EI95 & 1947--1989 (Annual) & 2005:Q1--2024:Q4 \\
        DP05 & 1970:Q1--2003:Q3 & same \\
        FHT25 & NA & 1981:M9--2025:M1 \\
        \bottomrule
    \end{tabularx}
    \vspace{0.3mm}\\    
    \noindent\textit{Notes:} See notes for Table \ref{tab:shock_studies} for the abbreviations used for studies. We report the original sample period and use ``same'' if replication codes were available and the original sample period was used for estimation. Otherwise, we give our extended estimation sample period. ``M'' and ``Q'' are used to indicate samples with monthly and quarterly data, respectively. The original sample in EI95 was based on annual data.
}
\end{table}

\subsection{Proxy Variables in Proxy VAR Studies}

Table \ref{tab:sw2012_shocks} provides the definitions and sources for the proxy variables used in our re-evaluation of the Stock and Watson (2012) application in Section 5.2. The first proxy variable in each category is used as the anchor proxy variable.

\begin{table}[htbp!]
    \caption{Summary of Structural Shocks, Proxies, and Definitions}
    \label{tab:sw2012_shocks}
    {\setstretch{1.25}\footnotesize
    \begin{tabularx}{\textwidth}{@{}
        >{\raggedright\arraybackslash}p{3.1cm}
        >{\raggedright\arraybackslash}X @{}}
    \toprule
    \textbf{Structural Shock} & \textbf{Proxy Definition and Source} \\
    \midrule

    \textbf{Oil}
    &\leavevmode\llap{$\ast$ }The percentage amount by which the oil price in a quarter exceeds the previous peak over the past 3 years \citep{hamilton2003oil}. \\
    & OPEC production shortfall stemming from wars and civil strife \citep{kilian2008exogenous}. \\
    & The residual from a regression of adjusted gasoline prices on various lagged macroeconomic variables \citep{ramey2011oil}. \\
    \midrule

    \textbf{Monetary policy}
    &\leavevmode\llap{$\ast$ }The residual of a constructed Federal Reserve monetary intentions measure regressed on internal Fed forecasts \citep{RomerRomer:2004}. \\[3mm]
    & The shock to the monetary policy reaction function in a DSGE model \citep{smets2007shocks}. \\[3mm]
    & The monetary policy shock from a structural VAR allowing for shifts in shock variances but constant VAR coefficients \citep{sims2006were}. \\[3mm]
    & The ``target'' factor, which measures surprise changes in the target federal funds rate \citep{GurkaynakSackSwanson:2005}. \\
    \midrule

    \textbf{Productivity}
    &\leavevmode\llap{$\ast$ }Quarterly total factor productivity adjusted for variations in factor utilization \citep{basu2006technology}. \\
    & The productivity shock in the Smets-Wouters DSGE model \citep{smets2007shocks}. \\
    \midrule

    \textbf{Uncertainty}
    &\leavevmode\llap{$\ast$ }The innovation in the VIX, computed as the residual from an AR(2) process \citep{bloom2009}. \\
    & The innovation in the common component of the economic policy uncertainty index \citep{baker2016measuring}. \\
    \midrule

    \textbf{Liquidity and }
     &\leavevmode\llap{$\ast$ }TED spread: Unadjusted term spread. \\
    \textbf{Financial Risk}  & A excess bond premium that has been adjusted to eliminate predictable default risk \citep{gilchrist2012}. \\
    & The unpredictable component of bank-level changes in lending standards \citep{bassett2014changes}. \\
    \midrule

    \textbf{Fiscal policy}
    &\leavevmode\llap{$\ast$ }Federal spending news instrument \citep{ramey2011fiscal}. \\
    & Measure of excess returns on stocks of military contractors \citep{fisherandpeters2010}. \\
    & Measure of tax changes relative to GDP \citep{romerandromertax}. \\

    \bottomrule
    \end{tabularx}    
    \vspace{0.3mm}\\    
    \noindent\textit{Notes:}
    Structural shocks and associated proxy variables in the Stock and Watson (2012) application. The table summarizes the proxy definitions and original sources for each shock category. An asterisk ($\ast$) marks the anchor proxy used to orient the sign of the composite instrument within each block.
    }
\end{table}

\end{document}